\documentclass[floatfix,12pt,onecolumn,nofootinbib,amsmath,amssymb]{revtex4-1}
\usepackage{color}
\usepackage{epsf}
\usepackage{graphicx}  % Include figure files
\usepackage{dcolumn}   % Align table columns on decimal point
\usepackage{bm}        % bold math
\usepackage[caption=false]{subfig}
\usepackage{xcolor}
\usepackage{amsmath}
\usepackage{amssymb}
\usepackage{amsfonts}
\newcommand{\be}{\begin{equation}}
\newcommand{\en}{\end{equation}}
 \newcommand{\bea}{\begin{eqnarray}}
 \newcommand{\ena}{\end{eqnarray}}

\begin{document}

\title{Effect of massive potentials on the holographic thermalization}
\author{Ali Nemati\footnote{Electronic address: ali.nemati@modares.ac.ir}, Shahrokh Parvizi\footnote{Corresponding author: Electronic address: parvizi@modares.ac.ir}}
\affiliation{Department of Physics, School of Sciences,
Tarbiat Modares University, P.O.Box $14155-4838$, Tehran, Iran }

%\date{ \today}

\begin{abstract}
We perform a numerical study to recognize the difference between various massive potentials in the dRGT massive gravity on the holographic thermalization in the AdS and AdS Gauss-Bonnet gravities. 
The massive potential in $4+1$ dimensions includes three symmetric polynomial terms which we denote them as $a_1$, $a_2$ and $a_3$ terms. 
We observe, in the case of time evolution of entanglement entropy that there is a critical size of the entangling surface on the boundary below which both signs of $a_1$ and above the critical size $a_3$ are able to reduce the thermal value of entanglement entropy. 
Our numerical computations show the more positive $a_i$'s are, the faster system reaches to its thermal value. The order of saturation time of positive potentials when supplemented to AdS or AdS-GB backgrounds is as $t_{sat}(a_{1})>t_{sat}(a_{2})>t_{sat}(a_{3})$.  We also explore these effects on the time evolution of the holographic mutual information.
\end{abstract}

%\pacs{04.20.-q, 04.70.-s}
%\keywords{Holographic thermalization, Vaidya solutions; Massive gravity; thermodynamics}

%\preprint{arXiv: }
 \maketitle

\section{Introduction}
The AdS/CFT correspondence~\cite{Maldacena:1997re,Gubser:1998bc,Witten:1998qj} provides a useful and powerful tool to study the details of strongly correlated quantum systems by postulating a duality with the dynamics of a classical gravity in one dimension higher. The idea has passed numerous tests. The AdS/CFT is the first realization of the holography principle, which is addressed that, all information of a gravitational system in a spatial region is encoded on the boundary of the region~\cite{tHooft:1993dmi,Susskind:1994vu}.
On the other hand, the entanglement entropy is an important and useful quantity in studying the quantum systems out of the equilibrium and the process of thermalization. For example, the thermalization time of quark-gluon plasma (QGP) produced in the ultra relativistic heavy-ion collisions at the Relativistic Heavy Ion Collider (RHIC) and the Large Hadron Collider (LHC), computed by the traditional methods of perturbation is longer than the experimental results~\cite{Gyulassy:2004zy}. This discrepancy reveals that we must find another powerful quantity and study its properties. One line of research is considering non-local probes such as two-point correlation function, Wilson loops and entanglement entropy. 

But what is the meaning of thermalization? The unitary time evolution of a system after a global quench~\cite{AbajoArrastia:2010yt} is considered as an example of thermalization. Due to limited methods of computing the non-local probes in  strongly correlated quantum systems, we should use the holography and AdS/CFT techniques. The idea of holographic quench~\cite{Hubeny:2007xt,AbajoArrastia:2010yt} is that the system is simply AdS at early times and after the inclusion of the quench as a shell of collapsing null dust, a black hole forms at late times. The thermalization process is studied widely in different backgrounds \cite{Fonda:2014ula,Albash:2010mv,Bai:2014tla,Zhang:2014cga,Li:2013cja,Caceres:2015bkr}. In the dual field theory the vacuum state of the system at early times is evolved by varying some tunable parameters of the Hamiltonian (e.g. the magnetic field) at some later time. 

To implement the holographic description, one may use Vaidya metrics \cite{Stephani:2003ika} which are the solutions of Einstein equations with negative cosmological constant and a nontrivial energy momentum tensor which describes the formation of the dual black hole through the collapsing of a shell of null dust. These metrics are widely used in the setup of realization of the holographic quench. 

In this article, we consider the thermalization process of a theory without momentum conservation in the boundary which in the dual background denoted by \textit{massive gravity}. In \cite{Blake:2013owa}, the existence of lattice (or inhomogeneity) on the boundary is related to the graviton mass in the bulk theory. Inspired by this, we perform a numerical study to realize which of terms of massive background have more effect on the thermalization process. In \cite{Hu:2016mym}, the effect of graviton mass is explored in 4-dimensional AdS bulk space-time on the thermalization process. We extend that setup by taking into account the higher curvature gravity corrections as Gauss-Bonnet coupling and study those effects in 5-dimensional AdS bulk. In fact, we evaluate the effect of both signs of massive potentials in the action as representatives (which was not considered in \cite{Hu:2016mym}). In this paper, we mainly focused on the qualitative behaviors of the effects appear in the course of thermalization.

The organization of this paper is as follows: In section \ref{sec:background}, we briefly review the action and Vaidya metric of AdS Gauss-Bonnet massive gravity in $d+1$ dimensional space-time. In section \ref{HEE}, the holographic entanglement entropy of a strip entangling region in the AdS-GB massive background is obtained. At the numerical section \ref{num1}, our findings are represented. Evolution of the holographic mutual information in a time-dependent AdS-GB massive background is also investigated in section \ref{MI} and finally section \ref{conclusion} is devoted to the conclusion and summary of results.

\section{Backgrounds with AdS-Vaidya Gauss-Bonnet massive gravity}\label{sec:background}
Consider the action of $d+1$ dimensional Gauss-Bonnet massive gravity with a negative cosmological constant $\Lambda$ as~\cite{Vegh:2013sk,Cai:2014znn,Xu:2015rfa}
\begin{equation}
\label{action}
I =\frac{1}{16\pi G_{N}^{d+1}}\int d^{d+1}x \sqrt{-g} \left[ R +2\Lambda+\lambda_{GB} L^2\, \mathcal{L}_{GB} +m^2 \sum^4_i c_i\, {\cal U}_i (g,f)\right],
\end{equation}
 where $m$ is the mass parameter, $L$ is the AdS length (hereafter for simplicity we will set $L=1$) and the other quantities are as follows
\begin{equation}
\Lambda =-\frac{d(d-1)}{2L^{2}}\quad,\quad \mathcal{L}_{GB}=\mathcal{R}^{2}-4\mathcal{R}_{{\mu}{\nu}}\mathcal{R}^{{\mu}{\nu}}+\mathcal{R}_{{\mu}{\nu}{\rho}{\sigma}}\mathcal{R}^{{\mu}{\nu}{\rho}{\sigma}}\,.
\end{equation}
In \eqref{action}, $c_i$'s are constants and $\mathcal{U}_i$ are symmetric polynomials of the eigenvalues of the $(d+1)\times (d+1)$ matrix $\mathcal{K}_{\nu}^{\mu}=\sqrt{g^{{\mu}{\alpha}}f_{{\alpha}{\nu}}}$,
\begin{eqnarray}
\mathcal{U}_{1} &=&\left[ \mathcal{K}\right] ,  \notag \\
\mathcal{U}_{2} &=&\left[ \mathcal{K}\right] ^{2}-\left[ \mathcal{K}^{2}%
\right] ,  \notag \\
\mathcal{U}_{3} &=&\left[ \mathcal{K}\right] ^{3}-3\left[ \mathcal{K}\right] %
\left[ \mathcal{K}^{2}\right] +2\left[ \mathcal{K}^{3}\right] ,  \notag \\
\mathcal{U}_{4} &=&\left[ \mathcal{K}\right] ^{4}-6\left[ \mathcal{K}^{2}%
\right] \left[ \mathcal{K}\right] ^{2}+8\left[ \mathcal{K}^{3}\right] \left[
\mathcal{K}\right] +3\left[ \mathcal{K}^{2}\right] ^{2}-6\left[ \mathcal{K}%
^{4}\right] ,  \notag
\label{potential}
\end{eqnarray}
and finally $f$ is a fixed symmetric tensor or reference metric which we will introduce soon. 
From the action (\ref{action}), we can derive the equations of motion by varying with respect to the metric tensor $g_{\mu \nu}$ 
\be
G_{\mu \nu}+\Lambda g_{\mu \nu}+\mathcal{H}_{\mu\nu}+m^{2}\mathcal{\chi}_{\mu\nu}=0
\label{eom}
\en
where $G_{\mu\nu}$ is the Einstein tensor, $\mathcal{H}_{\mu\nu}$ and $\mathcal{\chi}_{\mu\nu}$ are
\begin{eqnarray}
\mathcal{H}_{\mu \nu } &=&-\frac{\lambda_{GB} }{2}\left( 8R^{\rho \sigma }R_{\mu \rho \nu
\sigma }-4R_{\mu }^{\rho \sigma \alpha }R_{\nu \rho \sigma \alpha
}-4R_{\mu \nu }R+8R_{\mu \alpha }R_{\nu }^{\alpha }+g_{\mu \nu
}\mathcal{L}_{GB}\right) , \\
\mathcal{\chi} _{\mu \nu } &=&-\frac{c_{1}}{2}\left( \mathcal{U}_{1}g_{\mu \nu }-%
\mathcal{K}_{\mu \nu }\right) -\frac{c_{2}}{2}\left( \mathcal{U}_{2}g_{\mu
\nu }-2\,\mathcal{U}_{1}\mathcal{K}_{\mu \nu }+2\,\mathcal{K}_{\mu \nu
}^{2}\right) -\frac{c_{3}}{2}(\mathcal{U}_{3}g_{\mu \nu }-3\,\mathcal{U}_{2}%
\mathcal{K}_{\mu \nu }  \notag \\
&&+6\,\mathcal{U}_{1}\mathcal{K}_{\mu \nu }^{2}-6\,\mathcal{K}_{\mu \nu }^{3})-%
\frac{c_{4}}{2}(\mathcal{U}_{4}g_{\mu \nu }-4\,\mathcal{U}_{3}\mathcal{K}_{\mu
\nu }+12\,\mathcal{U}_{2}\mathcal{K}_{\mu \nu }^{2}-24\,\mathcal{U}_{1}\mathcal{K%
}_{\mu \nu }^{3}+24\,\mathcal{K}_{\mu \nu }^{4}).
\end{eqnarray}
The above e.o.m (\ref{eom}) has a static black hole solution 
\be
ds^{2}=-f(r)dt^{2}+\frac{1}{f(r)}dr^{2}+\frac{r^{2}}{\tilde{L}^2}h_{ij}dx_{i}dx_{j} 
\label{metric}
\en
where $h_{ij}$ is the metric on a $d-1$ dimensional space with constant curvature $(d-1)(d-2)k$ and different horizon topology $k=-1,0,+1$. In addition, $\tilde{L}$ is the effective AdS radius. In deriving the black hole solution in \eqref{metric} we take the reference metric $f$ simply as~\cite{Vegh:2013sk,Cai:2014znn,Xu:2015rfa}
\be
f_{\mu \nu}=\left(0,0,c_0^{2} h_{i j}\right)
\label{refmetric}
\en
where $c_0$ is a positive constant. Then it follows that 
\begin{eqnarray}
\mathcal{U}_1 &=& \frac{(d-1)c_0}{r}, \notag \\
\mathcal{U}_2 &=& \frac{(d-1)(d-2)c_{0}^{2}}{r^{2}},\notag \\
\mathcal{U}_3 &=& \frac{(d-1)(d-2)(d-3)c_{0}^{3}}{r^{3}} , \notag \\
\mathcal{U}_4 &=& \frac{(d-1)(d-2)(d-3)(d-4)c_{0}^{4}}{r^{4}}. 
\label{poten}
\end{eqnarray} 

The emblackening factor $f(r)$ is derived as follows~\cite{Hendi:2015pda} 
\begin{eqnarray}
f\left(r\right) &=& k+ \frac{r^{2}}{2\lambda_{GB} (d-3)(d-2)}  \left(1-\sqrt{1+4\lambda_{GB} (d-2)(d-3)
	\left[-1+\frac{M}{r^{d}}+\Gamma \right]}\right) \notag\\
\Gamma &=&m^{2}\left[ \frac{(d-2)(d-3)c_{0}^{4}c_{4}}{r^{4}}+\frac{%
(d-2)c_{0}^{3}c_{3}}{r^{3}}+\frac{c_{0}^{2}c_{2}}{r^{2}}+\frac{c_{0}c_{1}}{(d-1)r}%
\right] 
\end{eqnarray}
where $M$ is the black hole mass and $k$ set zero to have a black brane. This metric is static and useless in studying the characteristics of a thermalization process in the boundary theory. Instead, we need a Vaidya-like version of the above metric. To convert the static metric to a Vaidya one, we use the following coordinate transformation to dive into the Eddington-Finkelstein coordinates
\be
dv=dt+\frac{1}{f(r)} dr\,.
\en
After some manipulations, setting an inverse radius $z=\frac{\tilde{L}^{2}}{r}$ and promoting the black hole mass $M$ to a time-dependent one $M(v)$, the Vaidya type solution of metric (\ref{metric}) is obtained 
\be
ds^{2}=\frac{\tilde{L}^{2}}{z^{2}}\left(-f(z,v)d\nu^{2}-2d\nu dz+d\bar{x}^{2}\right)
\label{vaidya}
\en
where $\tilde{L}$ is the effective AdS radius, $\bar{x}$'s correspond to the spatial coordinates on the boundary and $f(z,v)$ is a function of massive parameters and GB coupling as follows
\begin{eqnarray}
f(z,v) &=&\frac{\tilde{L}^2 }{2(d-3)(d-2)\lambda_{GB}} 
\left(1-\sqrt{1+4(d-2)(d-3)\lambda_{GB}\left(1-M(v)\left(\frac{{L}^2}{z}\right)^{-d}-\tilde{\Gamma}\right)}\right) \notag\\
\notag\\
\tilde{\Gamma} &=& \frac{a_1 z}{(d-1)\tilde{L}^2}+\frac{a_2 z^2}{\tilde{L}^4}+\frac{a_3 (d-2)z^3}{\tilde{L}^6}+\frac{a_4 (d-3)(d-2)z^4}{\tilde{L}^8} 
\label{f(z,v)}
\end{eqnarray}
where $a_i=c_i c_0^i m^2 $. In this coordinate system, the boundary is located at $z\rightarrow 0$.
The above time-dependent mass $M(v)$ is chosen such that it interpolates between zero and finite value $M>0$ in a strictly increasing manner. Choosing this way, we have a background of pure AdS-GB massive gravity in early times $(v=-\infty)$ and after evolving system, an AdS-GB massive black hole forms at late times $(v=\infty)$. This profile is usually chosen to be~\cite{AbajoArrastia:2010yt}.
\be
M(v)=\frac{M}{2} \left(1+\tanh{\frac{v}{v_0}}\right)
\label{massprofile}
\en
where $v_0$ is the thickness of the infalling null shell and $M$ is the mass of the final black hole.

\section{Holographic entanglement entropy}\label{HEE}
In the static backgrounds where $M(v)=0$ or some constant $M$, the entanglement entropy $S_{A}=-tr(\rho_{A} \log {\rho_{A}})$ in the boundary theory holographically is obtained by a prescription proposed by Ryu-Takayanagi(RT) in  \cite{Ryu:2006bv,Ryu:2006ef}  as
\be
S_{A}=\frac{1}{4G_{N}^{d+1}} \min{[Area(\gamma_A)]} 
\en
where $G_N^{d+1}$ is the Newton constant for $d+1$ dimensional bulk and $\gamma_A$ is defined as the minimal surface extended in the bulk and anchored at the entangling region $A$ in the boundary such that $\partial{A}=\partial{\gamma_A}$ ($\gamma_A$ is homologous to $A$).
In the case of time-dependent geometries, the RT prescription has been generalized in \cite{Hubeny:2007xt} which is known as HRT recipe. The only difference is in the interpretation of the anchored surface $\gamma_A$ which here is obtained as an extremal surface.  In the following we explain it in details for d+1 dimensional Vaidya like metrics.

We want to compute the entanglement entropy of a spatial region A in the boundary theory. Let's assume A to be a $(d-1)$ dimensional rectangle such that $x^{1}\in \left(-\frac{\ell}{2},\frac{\ell}{2}\right)$ and the other coordinates $x^{2},\ldots,x^{d-1} \in (0,\ell_{\bot})$  at some fixed boundary time $t_b$. One can also assume $\ell \ll \ell_{\bot}$. According to the HRT prescription, the entanglement entropy of a region $\cal A$ is given by the extremal surface $\gamma_A$ which is conveniently parametrized by $v\equiv v(x^{1}) , z\equiv z(x^{1})$ and whose boundary coincides with the boundary of $\cal A$ at $z=0$. Since $x^{1}$ is the only relevant coordinate, from now on we denote it simply $x$. This dual surface extends all the way in the bulk with imposing the following boundary conditions
\be
z\left(-\frac{\ell}{2}\right)=z\left(\frac{\ell}{2}\right)=0\qquad,\qquad v\left(-\frac{\ell}{2}\right)=v\left(\frac{\ell}{2}\right)=t_b
\label{inicon}
\en
In the Einstein gravity the area of such an extremal dual surface gives simply the entanglement entropy but in case of corrections as higher curvatures this approach fails to apply. In such cases one must have the appropriate functional which is not known for a general higher curvature theory. Nonetheless, in the Gauss-Bonnet gravity the  correct functional of the EE in $d+1$ dimensional space-time is obtained in  \cite{deBoer:2011wk,Hung:2011xb}
\be
S_{EE}=\frac{1}{4G_N^{d+1}}\int_{\Sigma} d^{d-1}x \sqrt{h}\left(1+\lambda_{GB} L^{2} \mathcal{R}_{\Sigma}\right)+\frac{\lambda_{GB} L^{2}}{2G_N^{d+1}}\int_{\partial \Sigma} d^{d-2}x \sqrt{\sigma}\mathcal K
\label{GBmassive}
\en
where $h$ and $\sigma$ are the determinant of the induced metric on the extremal surface and its boundary. $\mathcal{K}$ is the trace of extrinsic curvature of the boundary of $\Sigma$. $L$ is the AdS length. In addition, $\mathcal{R}_{\Sigma}$ is the Ricci scalar corresponding to the induced metric on $\Sigma$. In fact, the last term is supplemented to provide a good variational principle in extremizing this functional. 
Now using the Vaidya metric (\ref{vaidya}) the induced metric on $\Sigma$ is given by
\begin{equation}
ds^{2}=\frac{\tilde{L}^2}{z^{2}}\left(\left(1-f(v,z)v'^{2}-2z'v'\right)dx^{2}+\sum_i^{d-2}d\bar{x}_i^{2}\right)
\end{equation}
where $\tilde{L}$, effective AdS radius, is obtained by limiting $z\rightarrow 0$ in the emblackening factor.
\[
\tilde{L}=\sqrt{\frac{2\lambda_{GB}(d-2)(d-3)}{1-\sqrt{1-4\lambda_{GB}(d-2)(d-3)}}}
\]
For sake of simplicity we have omitted the explicit $x$-dependence of functions $v(x)$ and $z(x)$. In above $'\equiv \frac{d}{dx}$ and we have chosen the origin such that the functions $v(x)$ and $z(x)$ are even.
For simplifying the computations in following we restrict to $d+1=5$ dimensional spacetime. Thus the required quantities for $S_{EE}$ is listed as follows 
\begin{eqnarray}
ds^{2}= \frac{\tilde{L}^2}{z^{2}}\left(\left(1-f(v,z)v'^{2}-2z'v'\right)dx^{2}+dx_2^{2}+dx_3^{2}\right) \\
\sqrt{h}=\frac{\tilde{L}^3}{z^{3}}\left(1-f(v,z)v'^{2}-2z'v'\right)^{\frac{1}{2}}\quad , \quad \sqrt{\sigma}=\frac{\tilde{L}^{2}}{z^{2}}, \notag\\
\sqrt{h} \mathcal{R}_{\Sigma} =-\frac{2\tilde{L} Q'z'}{Q^{\frac{3}{2}}z^{2}}-\frac{6\tilde{L} z'^{2}}{\sqrt{Q}z^{3}}+\frac{4\tilde{L} z''}{\sqrt{Q}z^{2}}, \notag
%\\ 
%\sqrt{h}\mathcal{R}_{\Sigma} =-\frac{2 \tilde{L} Q'z'}{Q^{\frac 32}z^{2}}-\frac{6\tilde{L}z'^{2}}{\sqrt{Q}z^{3}}+\frac{4\tilde{L}z''}{\sqrt{Q}z^{2}}, \notag
\end{eqnarray}
in which we have defined $Q(x)\equiv 1-f(v,z)v'^{2}-2z'v'$.
Defining a unit normal vector of the boundary $\partial \Sigma$ by $n^{a}$, one can evaluate the extrinsic curvature as
\[
 \mathcal{K}=\sigma^{ab}\nabla_{a} n_b \,.
\]
In our case, this unit normal vector is clearly along the $x^{1}\equiv x$ axes, thus
\[ 
n^{a}=\left(\frac{\tilde{L}}{z}\sqrt{Q},0,0\right)
\]
and the extrinsic curvature is obtained as 
\[
\mathcal{K}=\frac{2z'}{\tilde{L}\sqrt{Q}} \,.
\]
So the contribution of the Gibbons-Hawking term is such that it cancels with some of terms from intrinsic curvature of induced metric on the specific extremal surface
\[
\int{dx \frac{d}{dx}\left(\frac{-2\tilde{L}z'}{\sqrt{Q}z^{2}}\right)}  \,.
\]

Thus the entanglement entropy of a rectangular shape in the boundary is given by the following functional 
\be
S_{EE}=\frac{\ell^{2}_{\bot}}{4G_N^{5}}\int_{-\frac \ell 2}^{\frac \ell 2}\frac {1}{z^{3}}\left(\tilde{L}^{3}\left(1-f(z,v)v'^{2}-2z'v'\right)^{\frac 1 2}+\frac{2\lambda_{GB}\tilde{L}z'^{2}}{\left(1-f(z,v)v'^{2}-2z'v'\right)^{\frac 1 2}}\right)
\label{functional}
\en

As a check, in limit of $\lambda_{GB}\rightarrow 0$ one can obtain the results of entanglement entropy in absence of curvature corrections as area of an extremal surface\cite{Balasubramanian:2011ur}.
By extremizing the above functional, the equations of motion are obtained as follows, which is very dirty and cumbersome~
\begin{eqnarray}
z'' = \frac{P\left(z,z',v,v'\right)}{D\left(z,z',v,v'\right)} \notag\\
v'' = \frac{Q\left(z,z',v,v'\right)}{D\left(z,z',v,v'\right)}
\label{e.o.m}
\end{eqnarray}
where 
\begin{align}
P\left(z,z',v,v'\right) &= v'(x)^2 f^2 \bigg[-\tilde{L}^2 z(x) v'(x)^2 \partial_z f+12 z'(x) \left(2 \tilde{L}^2 v'(x)+ \lambda_{GB}  z'(x)\right)-12 \tilde{L}^2 \bigg]\notag\\ 
&- z(x) v'(x) \big(2 \tilde{L}^2 v'(x) z'(x)-\tilde{L}^2 +6 \lambda_{GB}  z'(x)^2 \big)\notag\\
&\big(v'(x) \partial_v f+2 z'(x) \partial_z f\big)+f \Bigg[v'(x)^2 \bigg(z(x) \left(\tilde{L}^2-6 \lambda_{GB}  z'(x)^2\right) \partial_z f\notag\\
&+ 24 \tilde{L}^2 z'(x)^2 \bigg)-4 \tilde{L}^2 z(x) v'(x)^3 z'(x) \partial_z f-\tilde{L}^2 z(x) v'(x)^4 \partial_v f\notag\\
&+ 24 v'(x) z'(x) \left(\lambda_{GB}  z'(x)^2 -\tilde{L}^2 \right)+6 \left(\tilde{L}^2 -2 \lambda_{GB}  z'(x)^2 \right)\Bigg]+6 \tilde{L}^2 v'(x)^4 f^3\notag
\end{align}

\begin{align}
Q\left(z,z',v,v'\right) &= 2 z(x) v'(x)^3 z'(x) \left(\tilde{L}^2 \partial_z f^{-4} \lambda_{GB} \partial_v f \right) \notag \\
&- v'(x)^2 \bigg[2 z'(x)^2 \big(5 \lambda_{GB}  z(x) \partial_z f+12 \tilde{L}^2 \big)+z(x) \left(\tilde{L}^2 \partial_z f-4 \lambda_{GB} \partial_v f\right)\bigg] \notag \\
&+ v'(x)^2 f \Bigg[-8 v'(x) z'(x) \left(\lambda_{GB}  z(x) \partial_z f+3 \tilde{L}^2 \right)+z(x) v'(x)^2 \left(\tilde{L}^2 \partial_z f-4 \lambda_{GB} \partial_v f\right) \notag \\
&+ 12 \left(\tilde{L}^2-\lambda_{GB}  z'(x)^2 \right)\Bigg]+8 v'(x) z'(x) \left(\lambda_{GB}  z(x) \partial_z f+3 L^2 -3 \lambda_{GB}  z'(x)^2 \right) \notag \\
&- 6 \tilde{L}^2 v'(x)^4 f^2 -6 \tilde{L}^2 +12 \lambda_{GB}  z'(x)^2 \notag
\end{align}

\begin{align}
D\left(z,z',v,v'\right) &= 2 z(x) \Bigg[2 v'(x) z'(x) \left(4 \lambda_{GB}  f+\tilde{L}^2\right)+ \left(v'(x)^2 f-1\right) \left(4 \lambda_{GB}  f+\tilde{L}^2\right)+6 \lambda_{GB}  z'(x)^2\Bigg] \notag
\end{align}

where $f\equiv f(z,v)$ defined in (\ref{f(z,v)}).
One should solve numerically these two E.O.M (\ref{e.o.m}) subject to the following initial conditions
\be
z(0)=z_t\qquad , \qquad z'(0)=v'(0)=0 \qquad, \qquad v(0)=v_t
\label{initialcond}
\en
Given the values of two free (initial) parameters ($z_t,v_t$), one can generate the profiles $z(x)$ and $v(x)$. Then the physical time $t_b$ is read from these numerical solution through boundary conditions $z(\frac \ell 2)=\epsilon$, $v(\frac \ell 2)=t_b$, where $\epsilon$, a UV cutoff, is introduced since the area functional above is divergent and needs to be regularized. The divergence comes from the fact that the volume of any asymptotically AdS space-time is infinite and the extremal surfaces which we considered reaches the boundary. 

By studying the same problem in the pure AdS-GB massive space-time ($r_h \rightarrow 0$) one can extract the divergence term and subtract it to obtain a finite area which is the main quantity we are interested in.

\section{Numerical Results}\label{num1}
In the following the results of our numerical approaches are presented in $d+1=5$ dimensional bulk space-time. In fact, first we study the effect of graviton mass (through different potentials $\mathcal{U}_i$) and GB coupling on the time evolution of the entanglement entropy then their saturation times are obtained. 

Since this problem has several free parameters, including the constants $c_i$'s, graviton mass $m$ and GB coupling $\lambda_{GB}$, we reduced some of them by going into the 5-dimensional space-time. Since setting $d=4$ in (\ref{poten}) the contribution of $\mathcal{U}_4$ goes away (the reason why we limited ourselves to this dimension). On the other hand, parameters in the graviton mass term appear in combinations of $a_i \equiv c_i c_0^i m^2$. Besides, we set the thickness of the shell $v_S=0.01$ (almost thin shell limit), the final black hole mass $M=1$ and the UV cutoff $\epsilon=0.01$. 

\begin{figure}[h]
\centering
\includegraphics[scale=0.6]{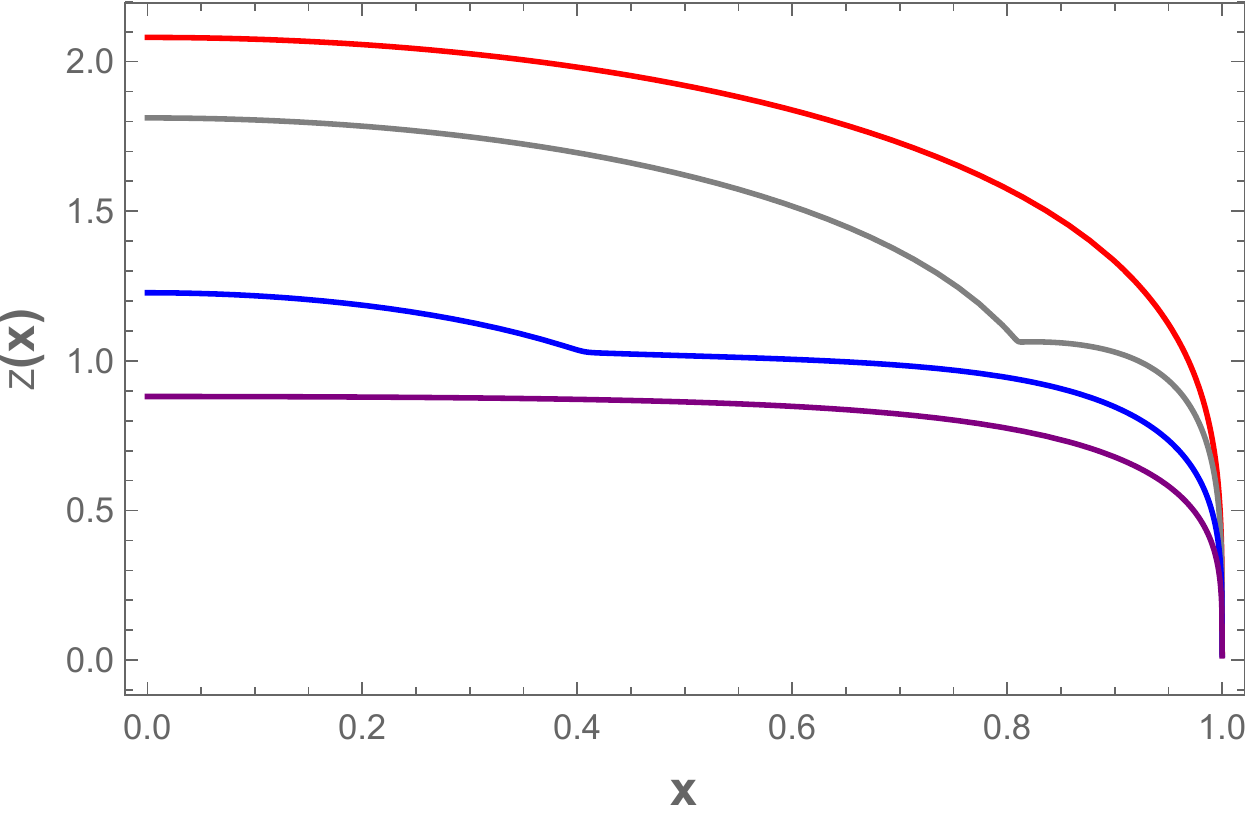}
\includegraphics[scale=0.6]{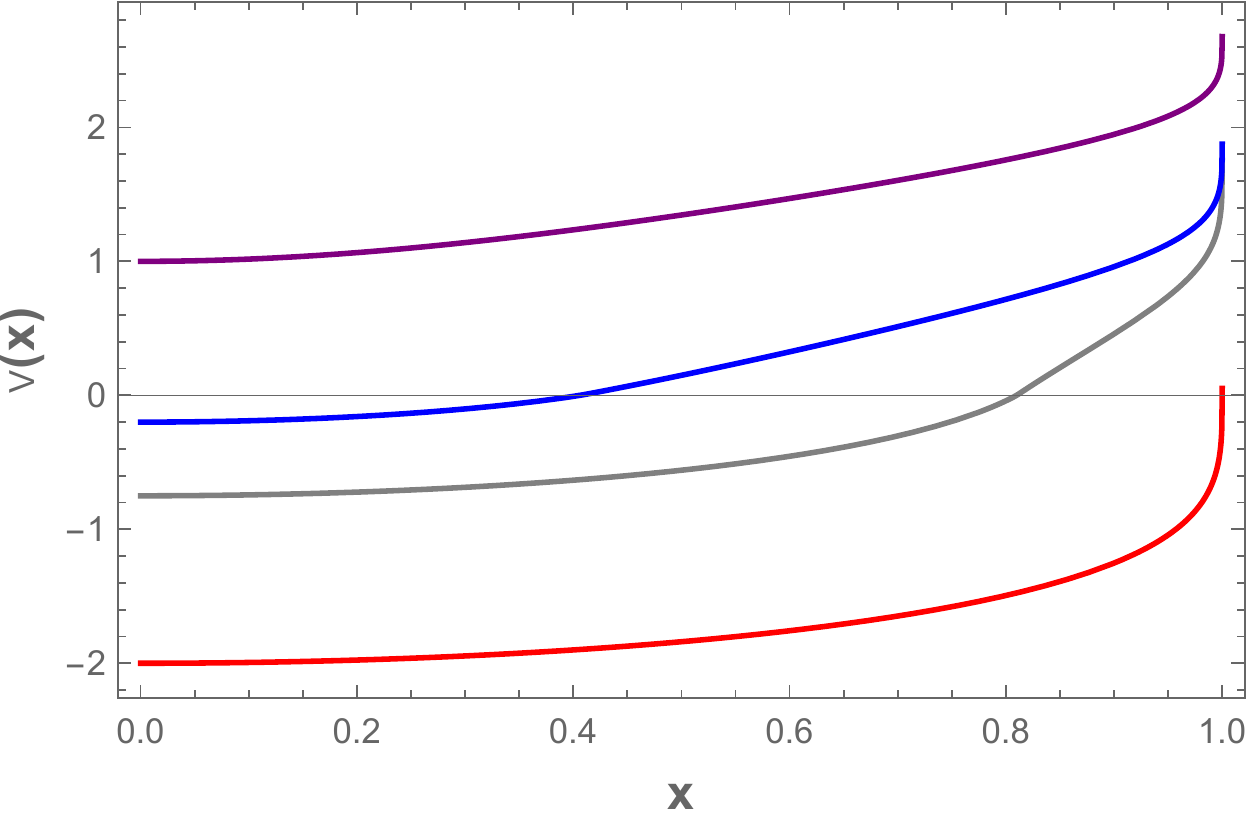}
\caption{\it The profile of extremal surfaces $z(x)$ for different boundary time $t$ in AdS-GB gravity without massive parameters i.e. $a_{1}=a_{2}=a_{3}=0$. The width of strip entangling region in the boundary theory is set $l=2$ and the Gauss-Bonnet coupling is $\lambda_{GB}=0.05$. The boundary times can be read from the right panel.}
\label{fig:surfplot1}
\end{figure}

In Fig \ref{fig:surfplot1}, the extremal surfaces for entangling surface $l=2$ is plotted. As one can see, once the boundary time $t_{phy} \leq 0$, the extremal surface is entirely in pure AdS-GB geometry, without any discontinuity feature (the red surfaces in $z(x)$ and $v(x)$). As times evolve and the black hole is forming, part of surfaces enters into the shell (the gray and blue lines) and for late times, when the black hole is completely formed, the extremal surface reaches the thermal equilibrium with its background and becomes independent of time (the purple  surface). 

\begin{figure}
	\centering
		\includegraphics[scale=0.8]{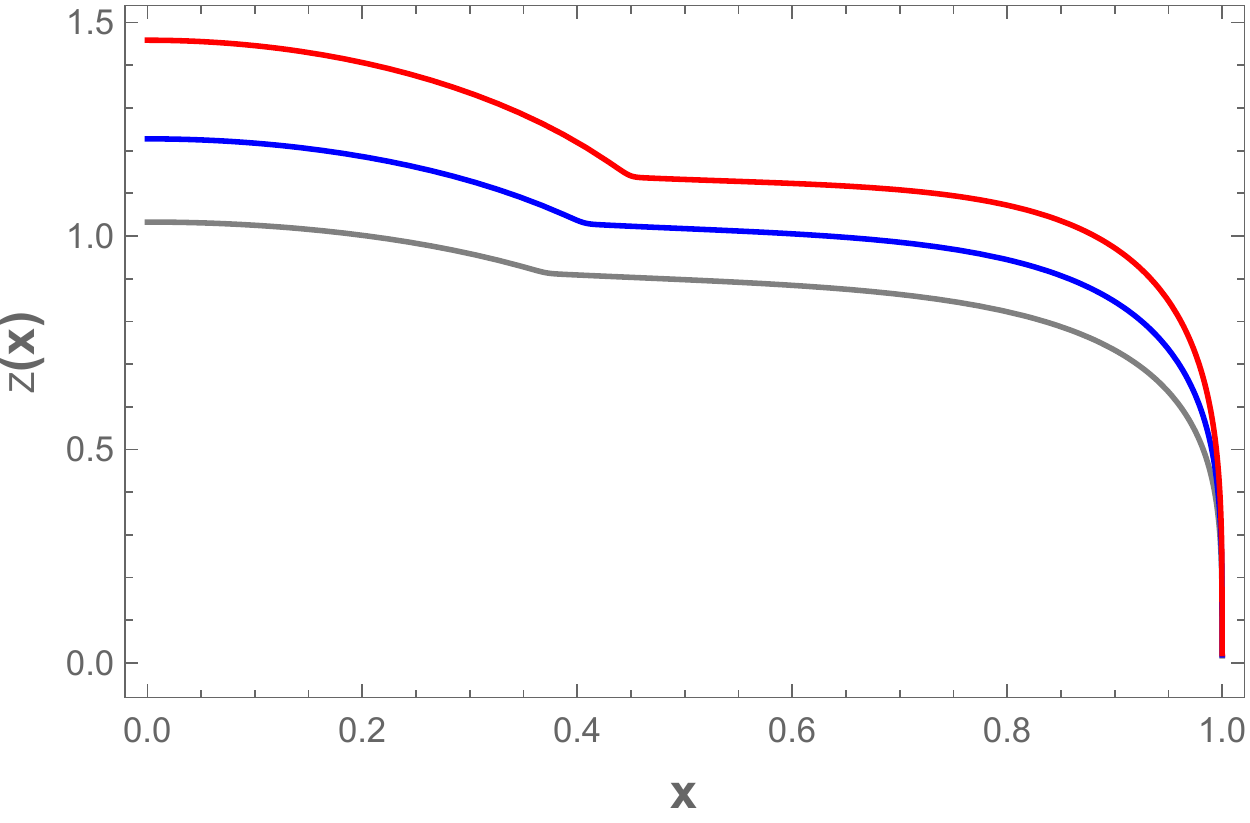}
	\caption{\it The effect of massive term $a_1$ on the extremal surfaces  for time boundary $t_{phy}\sim 1.9s$ in the AdS-GB massive gravity. Different color coding are $a_{1}=-1 ,0 ,1$ from bottom to top, respectively. The other parameters are $\lambda_{GB}=0.05 ,\, a_{3}=a_{2}=0$. The blue curve shows the AdS-GB gravity.}
	\label{fig:surfplot3}
\end{figure}

Now turning on the massive parameters can be led to some shift in the above generic properties of extremal surfaces depending on the sign of potentials. For example, each of the massive terms individually causes an upward (downward) shift in the values of dip point of extremal surface for $a_{i}>0\,(a_{i}<0)$ but with different strengths. In Fig \ref{fig:surfplot3}, this raising and falling feature of extremal surfaces is plotted for, say, the blue line in the Fig \ref{fig:surfplot1}.

To study the effect of massive terms on the evolution of entanglement entropy, one must define a renormalized one. One way to obtain a finite result is by subtracting the area of extremal surface at early or late times, once the black hole is absent or is formed, respectively. Here we choose the early time regularization scheme i.e. $\Delta S_{HEE}=S(t,\ell)-S_{vac}(\ell)$.
Now the aim is to study the effect of massive potentials and GB coupling on the time evolution of holographic entanglement entropy. The overall behavior of renormalized entanglement entropy in presence of these terms and corrections is the same as pure AdS. The only difference occurs in the final stage of time evolution where the black holes completely formed and EE computed in that background. That is the massive and GB corrections change the value of thermal EE. In the following, we plotted these effects for different signs of the coefficient of massive potentials with/without Gauss-Bonnet gravity as representatives\footnote{In \cite{Parvizi:2017boc}, authors discuss the causality condition on the boundary theory and show that mass parameters $a_i$'s may admit negative signs.}.

\begin{figure}[!h]
		\includegraphics[scale=0.6]{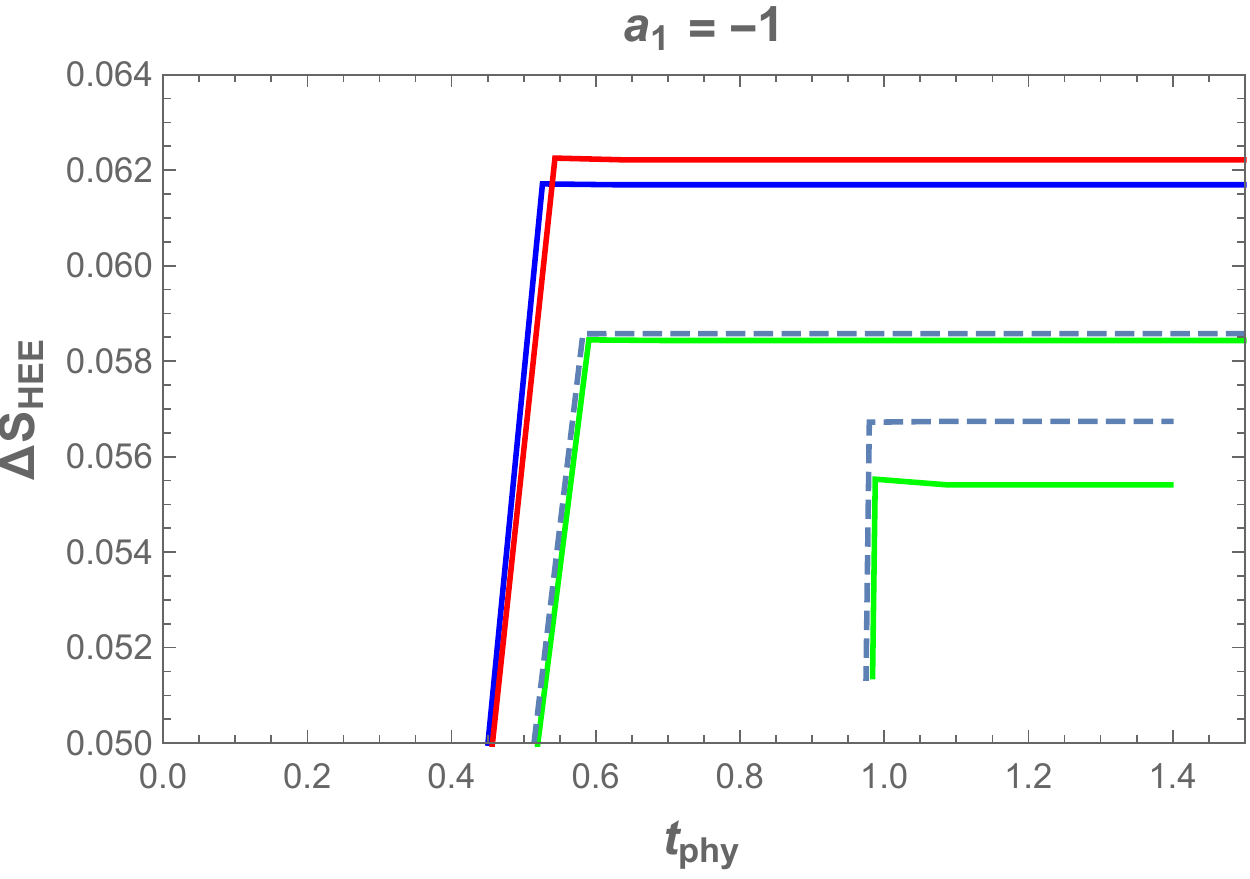}
		\includegraphics[scale=0.6]{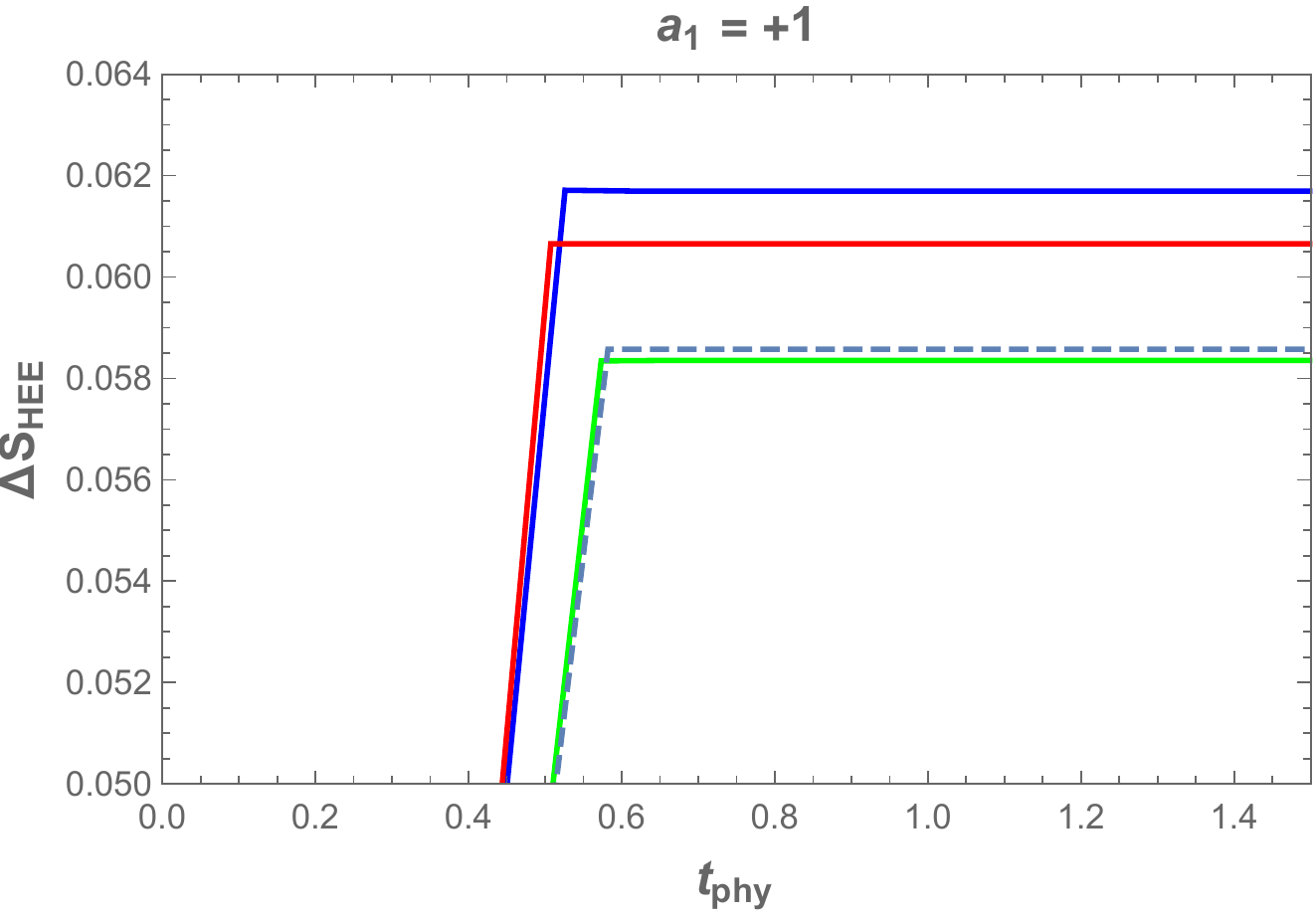}
		\includegraphics[scale=0.6]{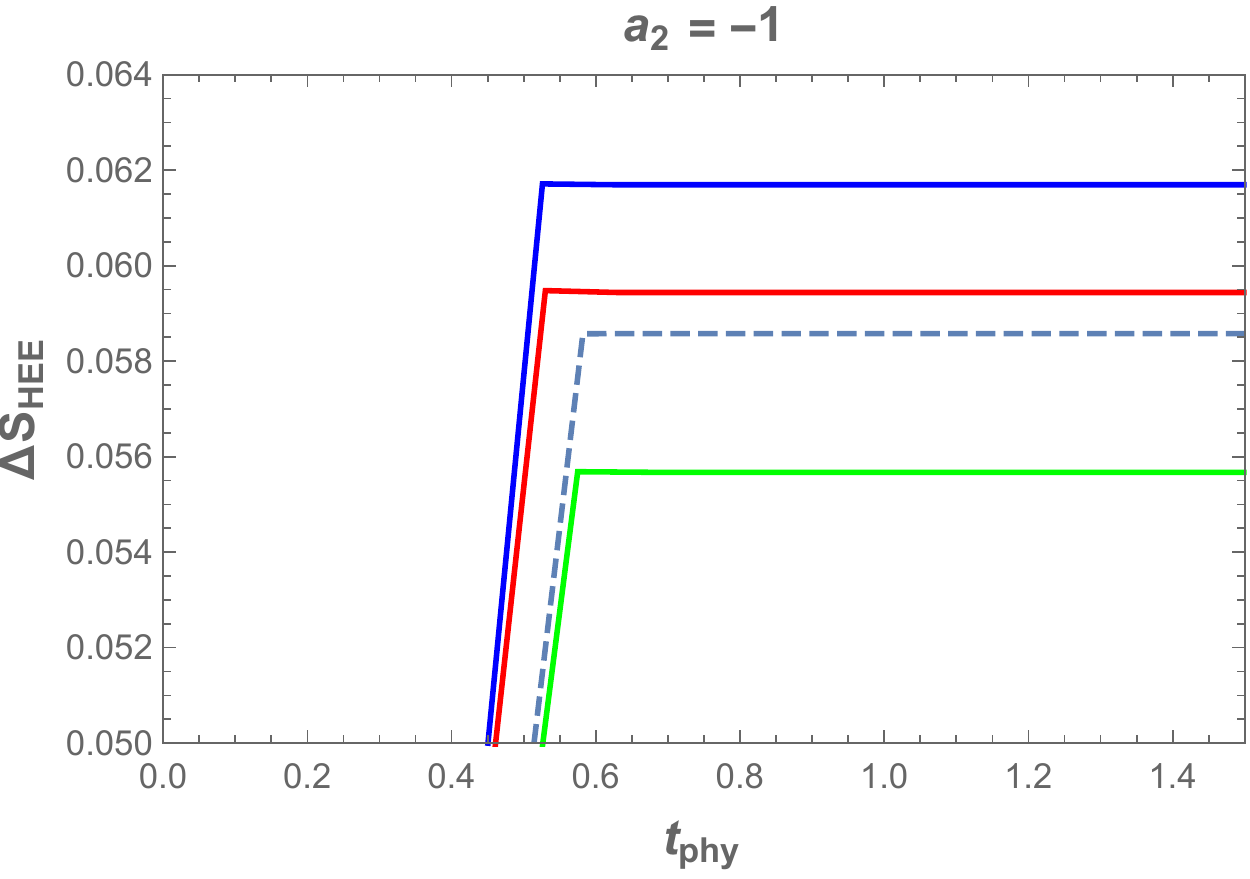}
		\includegraphics[scale=0.6]{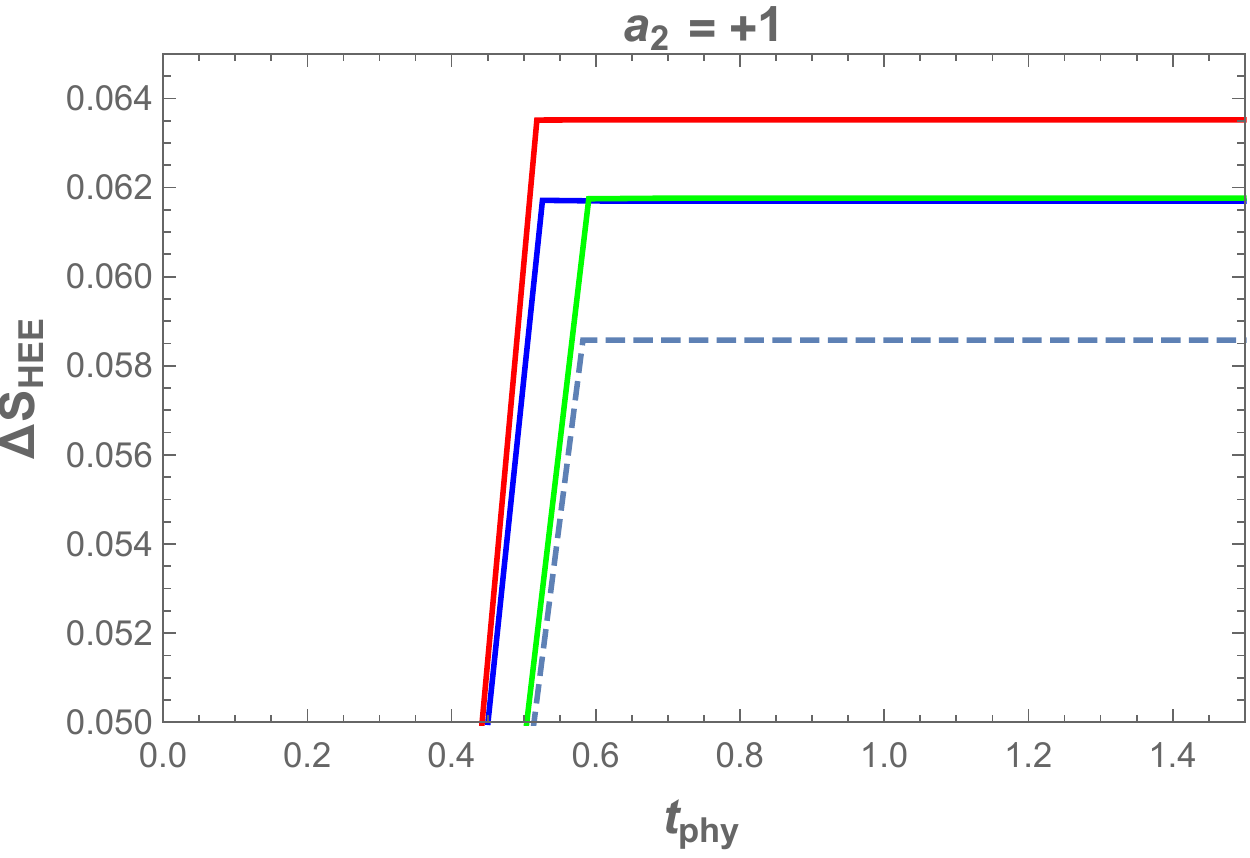}
		\includegraphics[scale=0.6]{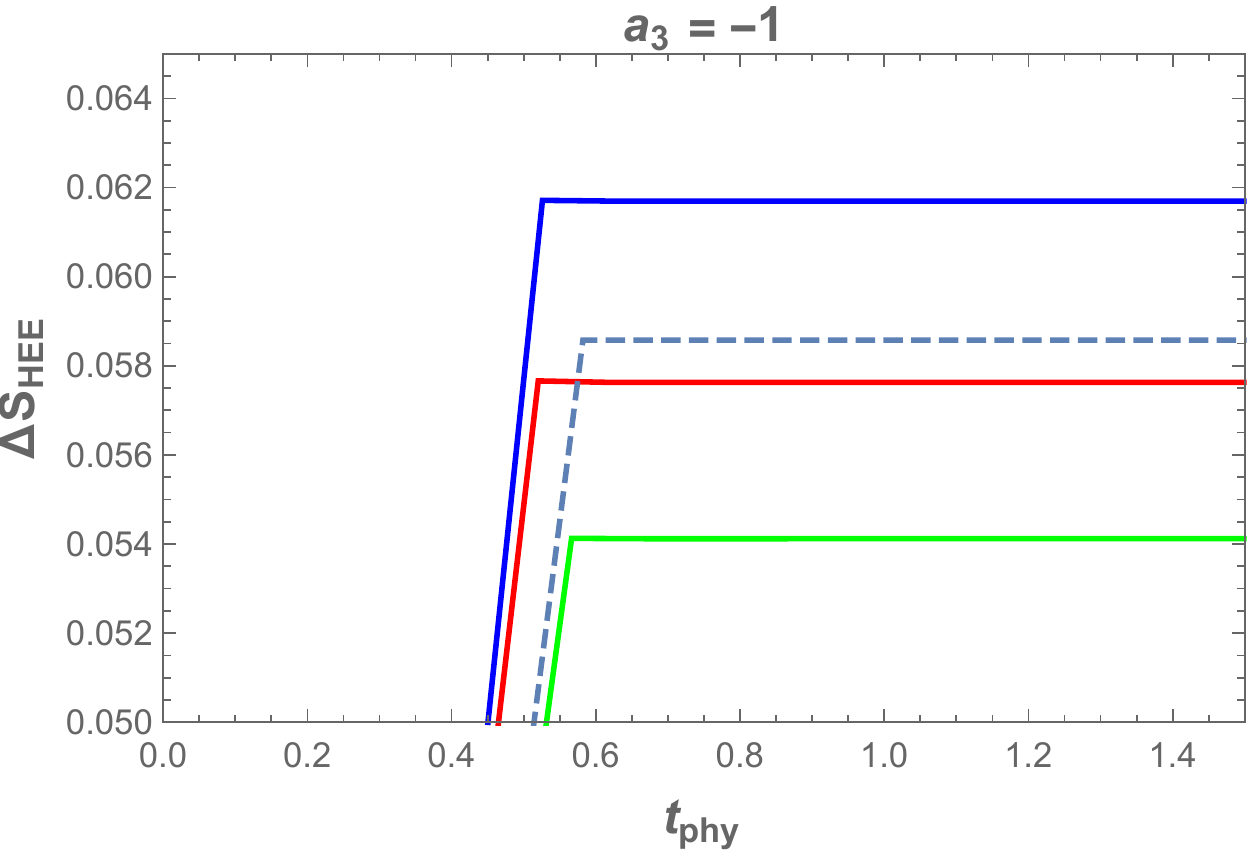}
		\includegraphics[scale=0.6]{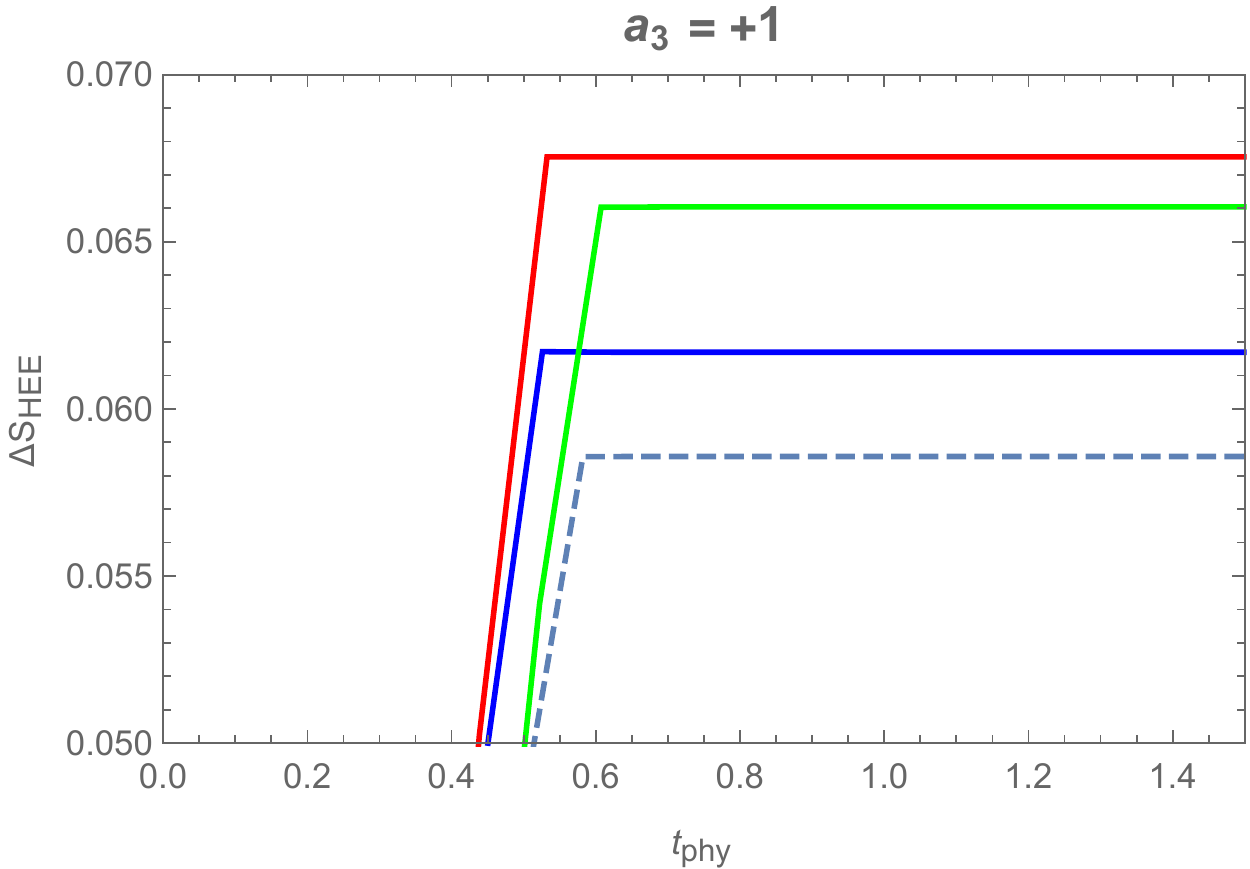}
\caption{\it Time evolution of renormalized entanglement entropy of a strip boundary region in different backgrounds including AdS-massive and AdS-GB massive gravities (only the thermal stage of evolution is shown). The blue, green and red lines represents the pure AdS-GB , AdS massive gravity without GB coupling and AdS massive gravity in presence of GB coupling, respectively. In all plots, the strip width and GB coupling are set as $\ell=0.5$ , $\lambda_{GB}=0.05$. The dashed gray line is for pure AdS background. The inset graph for $a_1<0$ shows the small variation in pure AdS background once it added.}
\label{fig:3}
\end{figure}

Before we dive in the details, it should be noted that depending on the choice of parameter space, there is a critical entangling interval which beyond that the only notable behavior is related to the $a_3$ term. Here, to avoid confusion, we only plotted for $\ell \leq \ell_{cr}$ in Fig \ref{fig:3} and zoomed in the final stage of the evolution. In these plots, there are two sets of drawing, one is for AdS (massive) backgrounds depicted by dashed (green) line and the other is for AdS-GB (massive) gravities which is shown by blue (red) line in Fig \ref{fig:3}. As one can see, there is no general rule about the behavior of different potentials except in case of $a_1$ in  AdS massive background (recall that beyond the $\ell_{cr}$ the $a_3$ term has such behavior).  Both signs of $a_1$ very slightly decreases the thermal value of HEE with respect to the pure AdS space-time. Besides, in both pure AdS and AdS-GB backgrounds the effect of massive terms in either of conditions $a_1>0$ , $a_3<0$ or $a_2<0$ is lowering the final value of HEE. Whereas for either of $a_1<0$ , $a_2>0$ or $a_3>0$, the presence of massive term leads to increase of HEE with respect to massless theory (except for $a_1$ in case of AdS-massive, mentioned above). 

\begin{figure}
	\centering
		\includegraphics[scale=0.6]{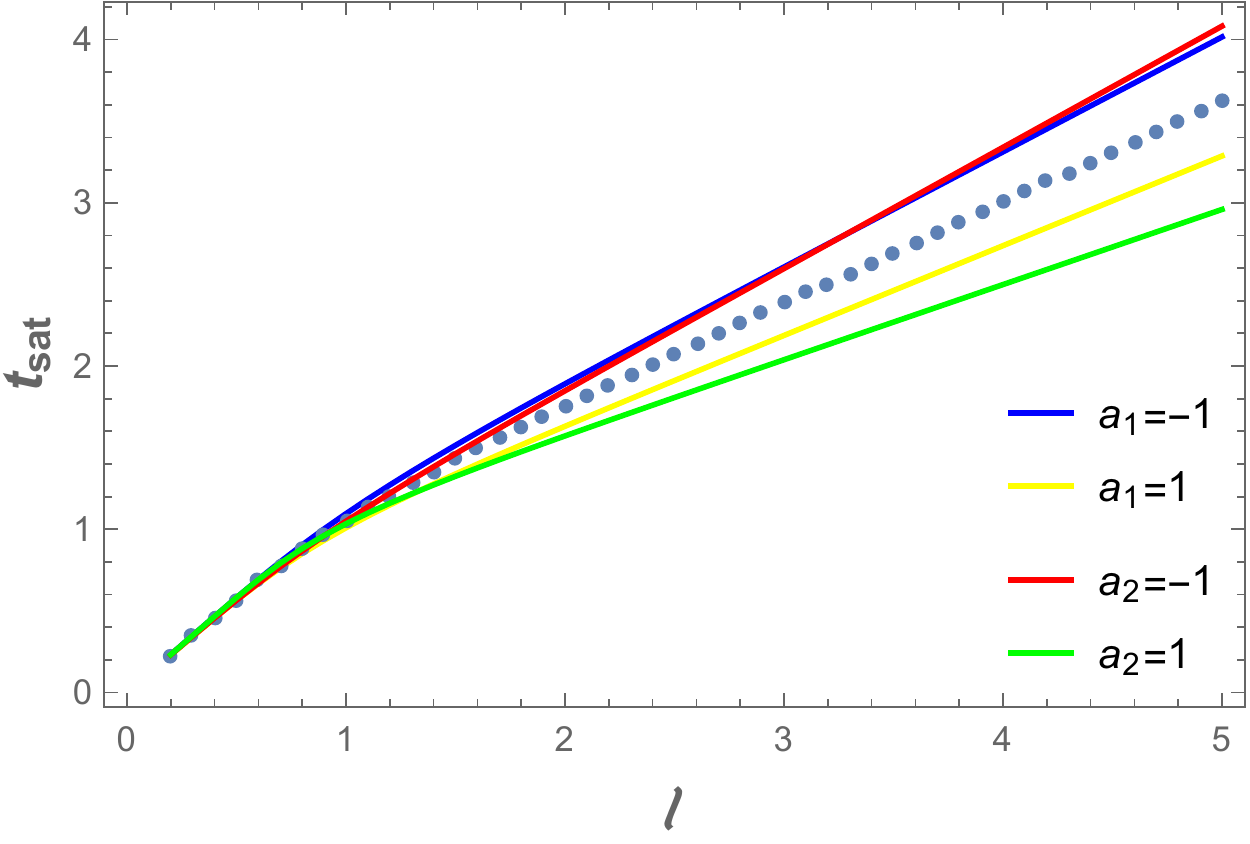}
		\includegraphics[scale=0.6]{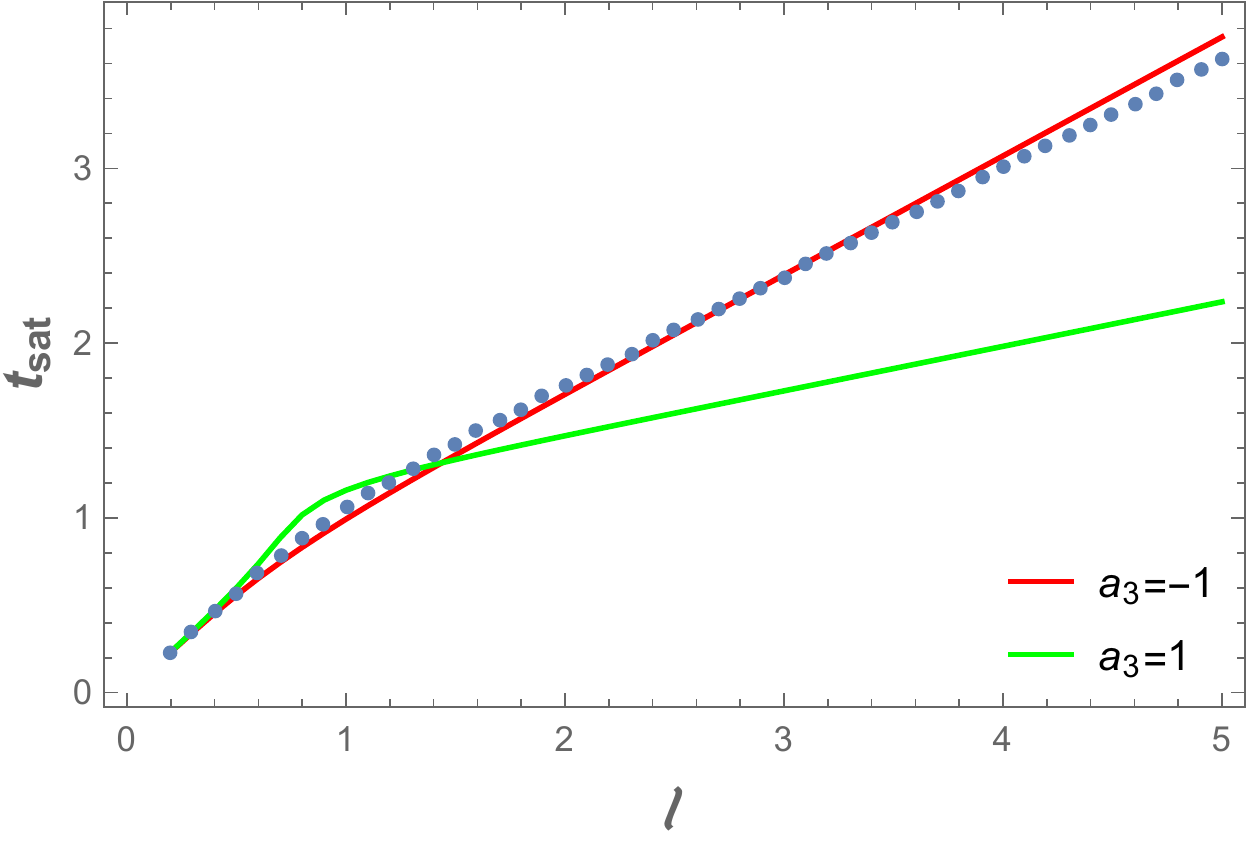}
	\caption{\it Saturation time $t_{sat}$ against width of boundary interval $\ell$ is plotted for various signs of massive parameters in AdS massive gravity. In plots, dotted curve represent the pure AdS in $d+1=5$ dimensional space-time. At very small widths, saturation time is independent of type and sign of massive term.}
\label{fig:tsat1}
\end{figure}

For completeness, we plotted the saturation time $t_{sat}$ against the width of the interval $\ell$ for both signs of $a_i$'s as representative. From these plots we can further learn that the saturation time for small boundary interval regions is independent of the choice of the massive parameters but at large intervals the situation is some different. The positive values of $a_1$ and $a_2$ lower the saturation time while their negative contributions increase it (Fig \ref{fig:tsat1}). On the other hand, in presence of the $a_2>0$ potential the system can thermalize faster than the $a_1>0$. In case of $a_3$ potential, there is a strange behavior. At sufficiently small entangling regions, the saturation time is independent of the sign of $a_3$ like before. The negative values of $a_3$ have \textit{almost} the same saturation time as the pure AdS (recall that the $a_i<0$ where $i=1,2$ increases the saturation time) but the positive $a_3$ as a function of $\ell$, first increases and then decreases the saturation time. 

\begin{figure}
	\centering
		\includegraphics[scale=0.6]{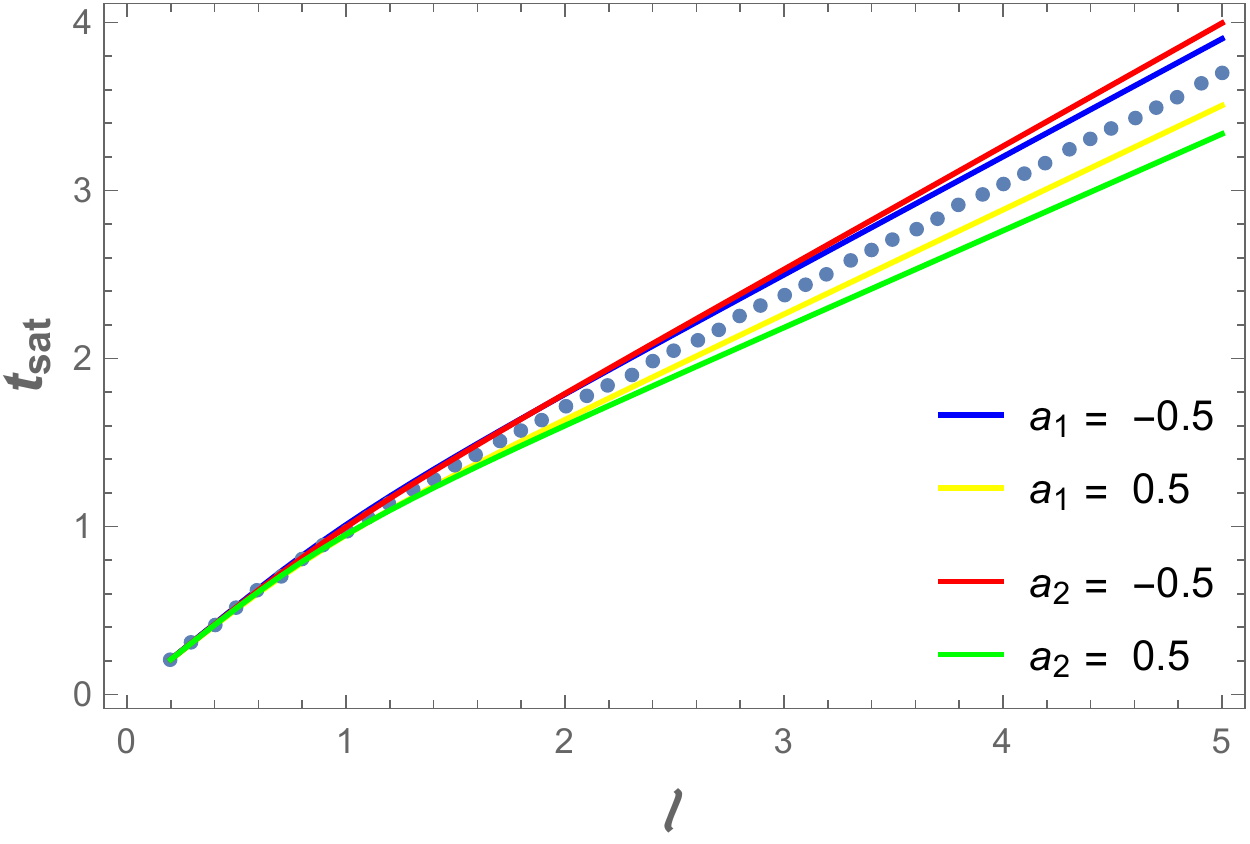}
		\includegraphics[scale=0.6]{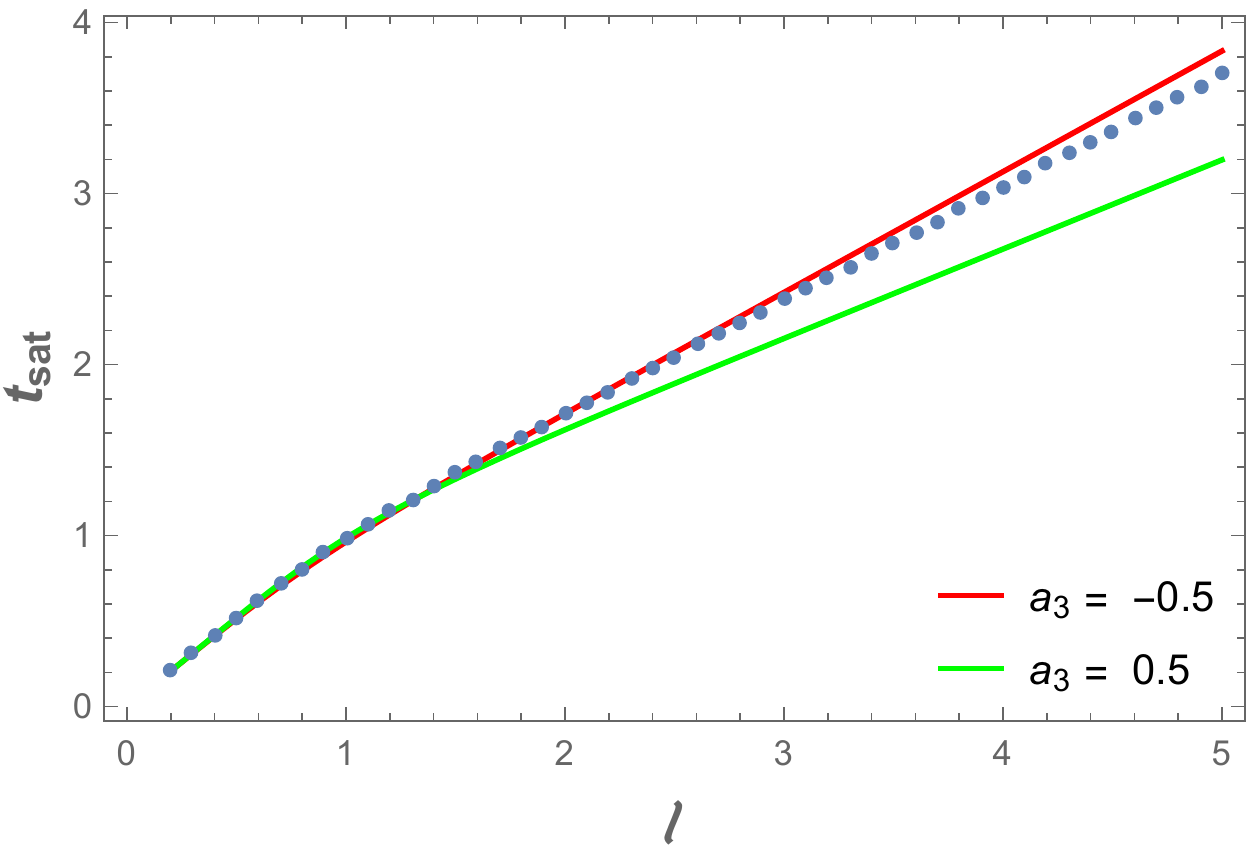}
	\caption{\it Saturation time against the width of boundary entangling region $\ell$ for either signs of massive potentials $a_i$ in the AdS-GB massive gravity. The GB coupling is set $\lambda_{GB}=0.05$ and the dotted curve represents the pure AdS-GB massive gravity. In the case of $a_{3}>0$, there is no raising in the time saturation at very small lengths.}
\label{fig:tsatgb1}
\end{figure}

In general, at large regions the system is thermalized faster by turning on the $a_i>0$ potentials as in the following order:
\[t_{sat}(a_{3})<t_{sat}(a_2)<t_{sat}(a_1)\]
where for comparison, we have supposed that all $a_i$'s values to be equal.\footnote{Notice that $a_i$'s have different dimensions. Nonetheless, we can firstly make them dimensionless by an appropriate factor of the AdS length which we have taken to be $L=1$. Then we consider equal values of $a_i$'s in the above inequalities.}  
Turning on the massive potentials in the AdS-GB background, does not make any significant change to the generic pattern of saturation time at small and large boundary regions for $a_1$ and $a_2$. In the case of $a_3>0$, the strange behavior seen in pure AdS (existence of a transition point at intermediate lengths) is more difficult to reveal and seen as a very tiny effect at small entangling regions (see Fig \ref{fig:tsatgb1}). 

\begin{figure}
	\centering
		\includegraphics[scale=0.7]{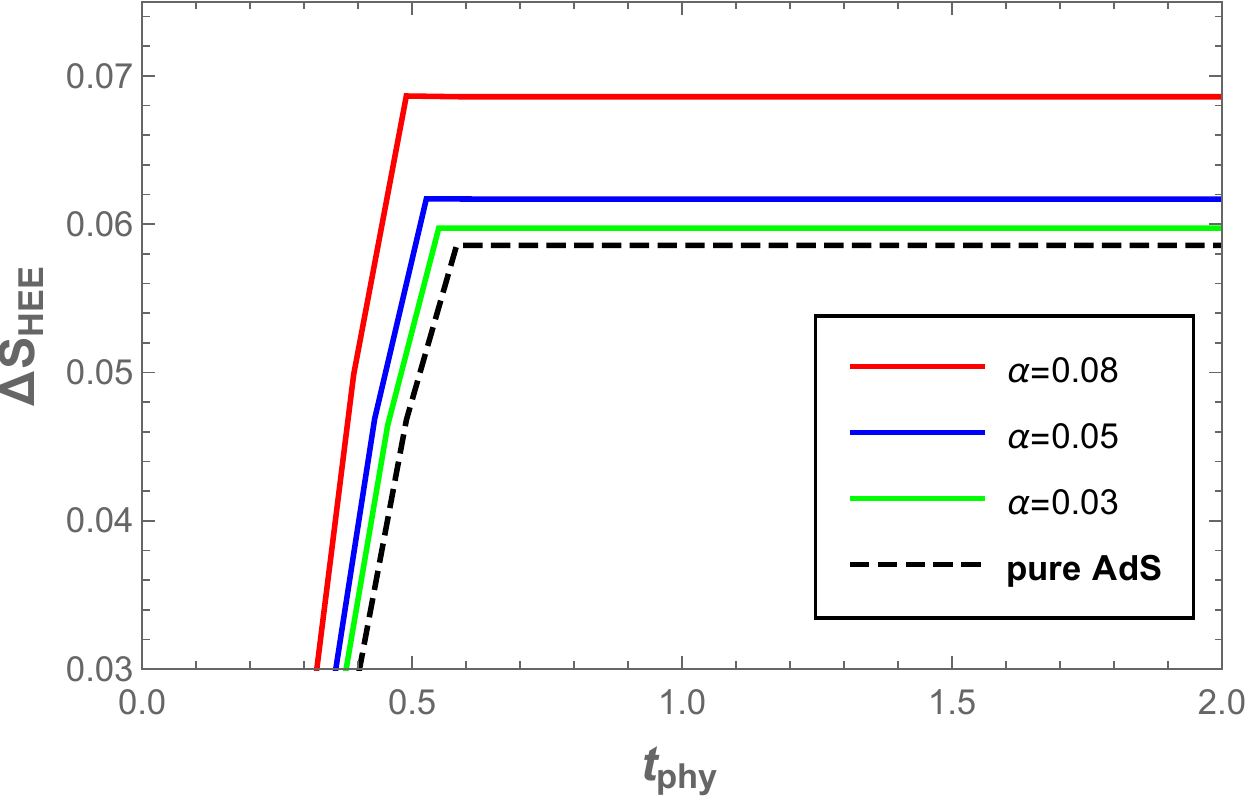}
		\caption{\it Effect of Gauss-Bonnet coupling on the renormalized EE in a time dependent background}
\label{fig:7}
\end{figure}

Another comment is about the effect of Gauss-Bonnet gravity individually. Turning on and increasing the GB coupling can be caused of more increase in the magnitude of equilibrated (thermal) EE (Fig \ref{fig:7}).
  By increasing the GB coupling, the EE reaches its thermal value sooner. This time which is called the saturation or thermalization time is reported for various couplings in the table \ref{table1}. One can observe that the saturation time decreases as $\lambda_{GB}$ increases.
\begin{table}
\begin{center}
\begin{tabular}{ | m{8em} | c | c | c | c | c | }
\hline
$\lambda_{GB}$  & $0.01$ & $0.03$ & $0.05$ & $0.07$ & $0.09$  \\
\hline
$\text{Saturation time}$  & 0.56203  &  0.53969  &  0.51659  &  0.49217  & 0.46561 \\
\hline
\end{tabular}
\caption{\it Effect of GB coupling on the thermalization time of renormalized EE in AdS-GB gravity. The width of strip in the boundary is $\ell =0.5$}
\label{table1}
\end{center}
\end{table}

\begin{figure}
	\centering
		\includegraphics{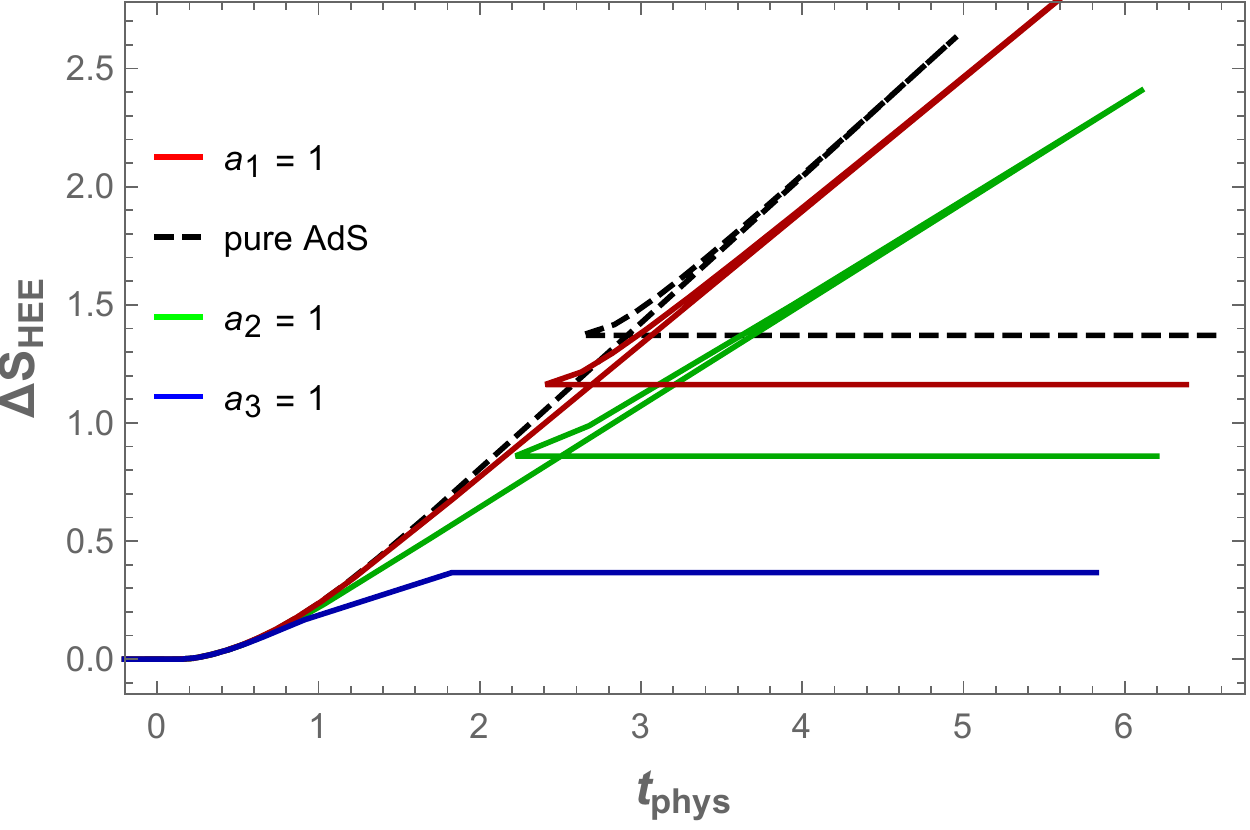}
		\includegraphics{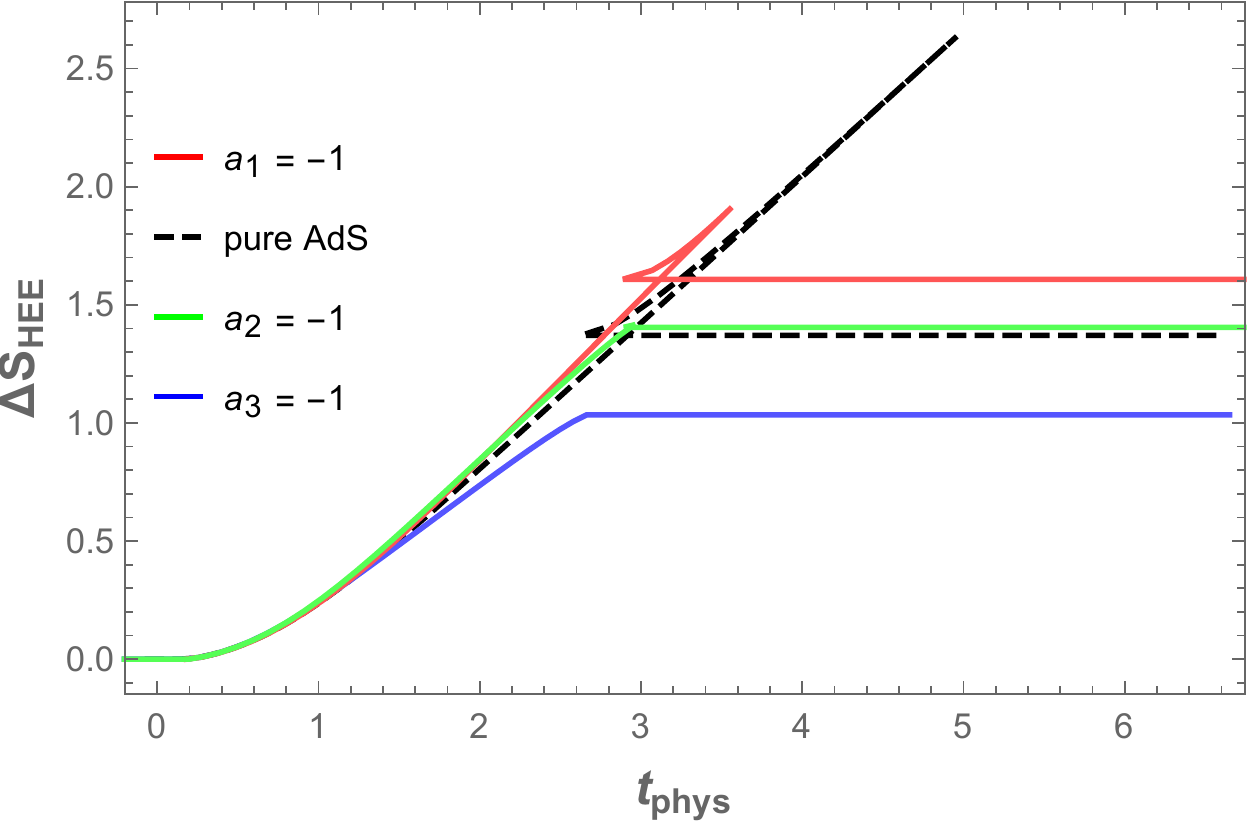}
	\caption{\it Effect of massive terms on the size of the kinks in $d+1=5$ dimensional AdS space-time. The width of strip entangling region is set $\ell=3.4$ where the kink appears.}
\label{fig:kink1}
\end{figure}

As a final note, the effect of massive terms on the situation of the kinks in AdS massive background is plotted. Recall that kink is appeared when the length of entangling region is large~\cite{Albash:2010mv}. In these cases, there are several extremal surfaces with different initial conditions which anchored at the same final point on the boundary.
As one can observe in Fig \ref{fig:kink1}, the large kink which is appeared in the large interval boundary $\ell=3.4$ in pure AdS, can be get smaller in size by turning on the negative massive potentials ($a_i<0$) but with different strengths in order: $a_{3}>a_{2}>a_{1}$. The positive coefficient of massive terms ($a_i>0$) does not have any significant effect on the size of the kink (in fact slightly enlarges) except the $a_3$ term which here, as before (i.e. $a_3<0$), completely removes the kink and also much lowers the value of thermal HEE. 

\begin{figure}
	\centering
		\includegraphics{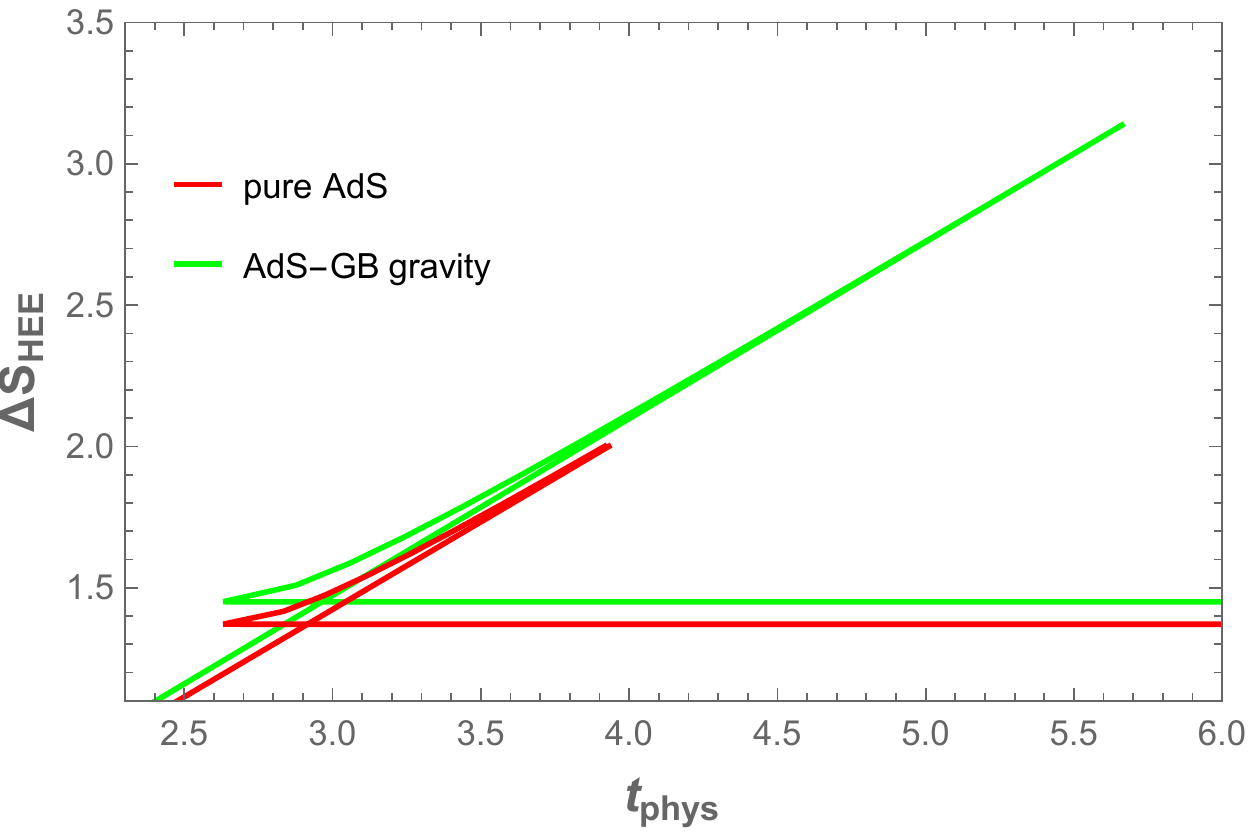}
	\caption{\it Effect of Gauss-Bonnet correction on the kink which is appeared in pure AdS. The width of entangling region is $\ell=3.4$ and $\lambda_{GB}=0.02$. In this plot, for more illumination, only the kink's part of the time evolution of entanglement entropy is shown.}
	\label{fig:kink3}
\end{figure}

It is worth mentioning that Gauss-Bonnet corrections can be able to enlarge the kink existed in pure AdS background (see Fig \ref{fig:kink3}).

Unfortunately, due to the limitation of numerical computations, we could not explore the effect of massive terms on the kinks in an AdS-GB background. We hope to study this issue in future.

\section{Mutual Information}\label{MI}
In this section we analyze the holographic mutual information for AdS Vaidya-GB massive gravity in $d=4+1$ dimensional bulk. In all findings, there is a time independent region where the holographic mutual information vanishes everywhere. 
Let's consider two separate spatial regions $A$ and $B$ in the boundary. By these, one can define a UV finite quantity as 
\be
I(A,B)=S(A)+S(B)-S(A\cup B)
\en
It is worth mentioning that when those two regions have a shared boundary, the holographic mutual information failed to be a finite quantity. So not having an overlapping boundary is the necessary condition to have a finite mutual information. In addition, it is always non-negative which can be seen in the following. 
The only subtlety of mutual information is computing the entanglement entropy of the union of two entangling regions. In fact, given two disconnected intervals in the boundary, there are three configurations characterizing the extremal surfaces extending in the bulk whose boundaries coincide with $\partial A=\partial{A}\cup \partial{B}$. The first is simply the union of two surfaces implying $S_A$ and $S_B$. This scheme referred to as \textit{disconnected} configuration. The second one connecting the initial and end points of one to the other which is referred as \textit{connecting} configuration. The third one which is called \textit{mixed} connecting the initial (end) points of one region to the other. Here, the entanglement entropy is an increasing function of the size of entangling region so we can claim that this mixed state is always suboptimal with respect to the disconnected one. In \cite{Headrick:2010zt}, it is shown that for equilibrium configurations, $S(A\cup B)$ is given by the minimum of two sets of extremal surfaces mentioned above. We will assume this rule is extended to the Vaidya metrics. Therefore, we have
\be
I(A,B)=S(A)+S(B)-\min\,(S_{con},S_{dis})
\en
where $S_{dis}$ and $S_{con}$ are disconnected and connected configurations.
When the distance between the two regions exceed from a specific point, the mutual information becomes zero. In other words, in the context of holographic CFTs there is a transition point in the configuration space of holographic mutual information which is not observed in the simple CFTs~\cite{Headrick:2010zt,Calabrese:2009ez,Calabrese:2010he,Tonni:2010pv}. This happens once the contributions of connected and disconnected configurations are equal.

\subsection{Numerical Results for the Mutual Information}
\begin{figure}[t]
	\centering
		\includegraphics[scale=0.6]{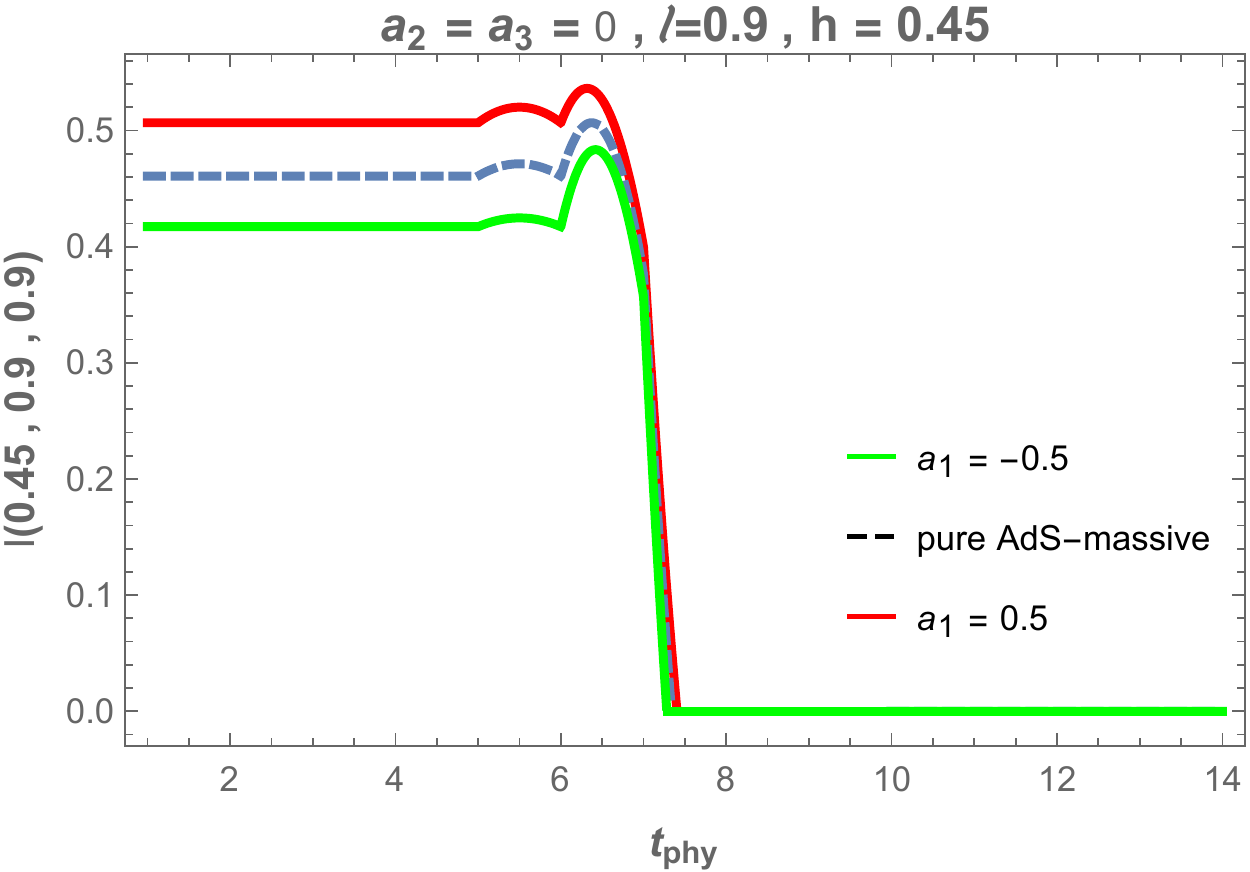}
		\includegraphics[scale=0.6]{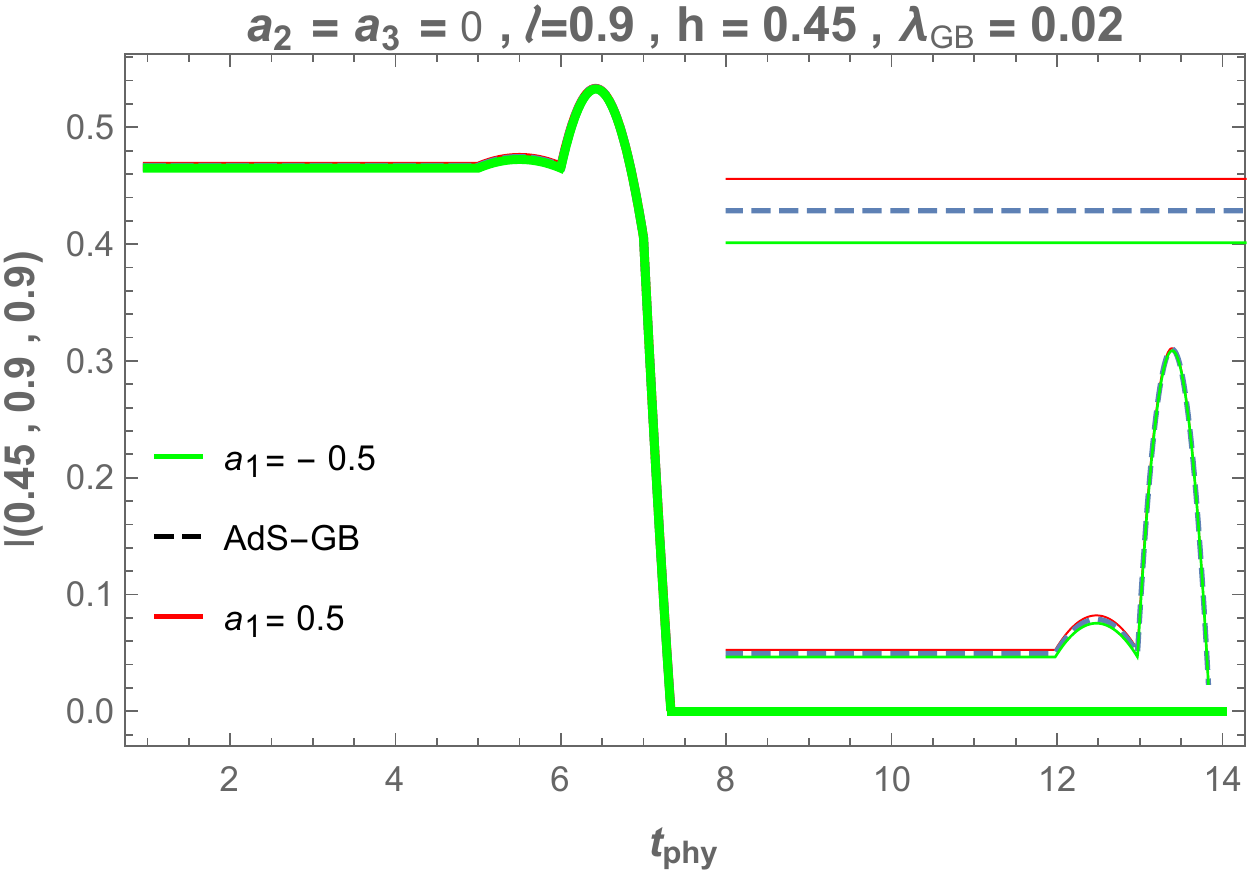}
		\includegraphics[scale=0.6]{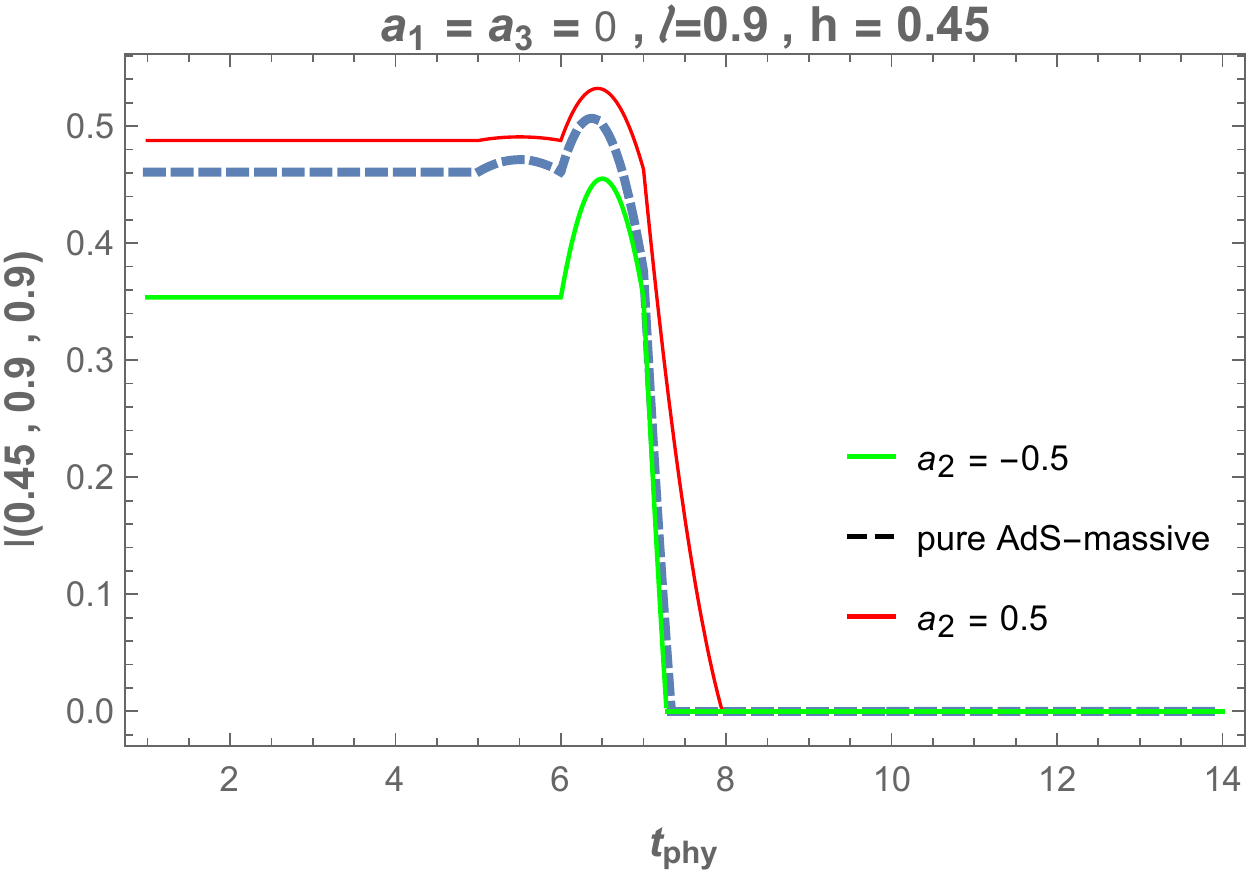}
		\includegraphics[scale=0.6]{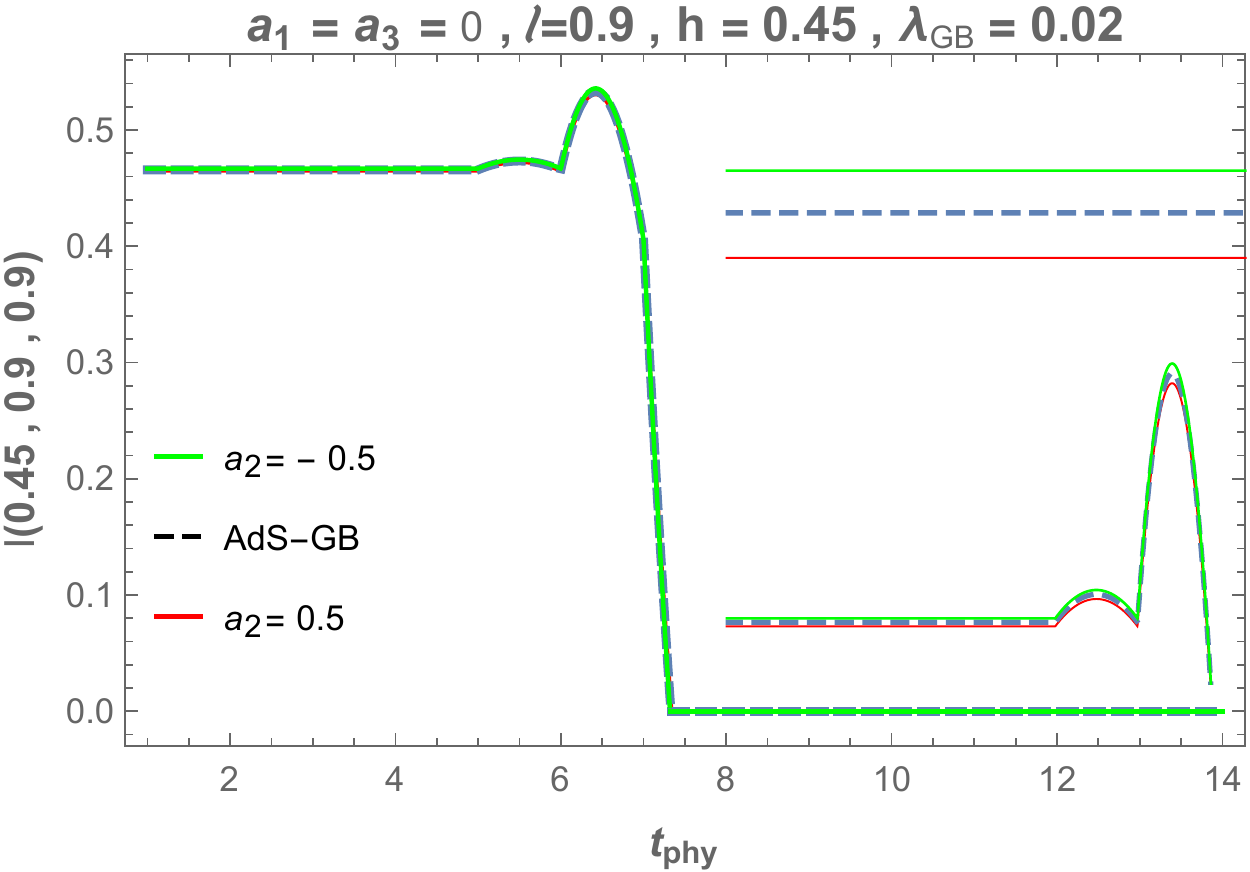}
		\includegraphics[scale=0.6]{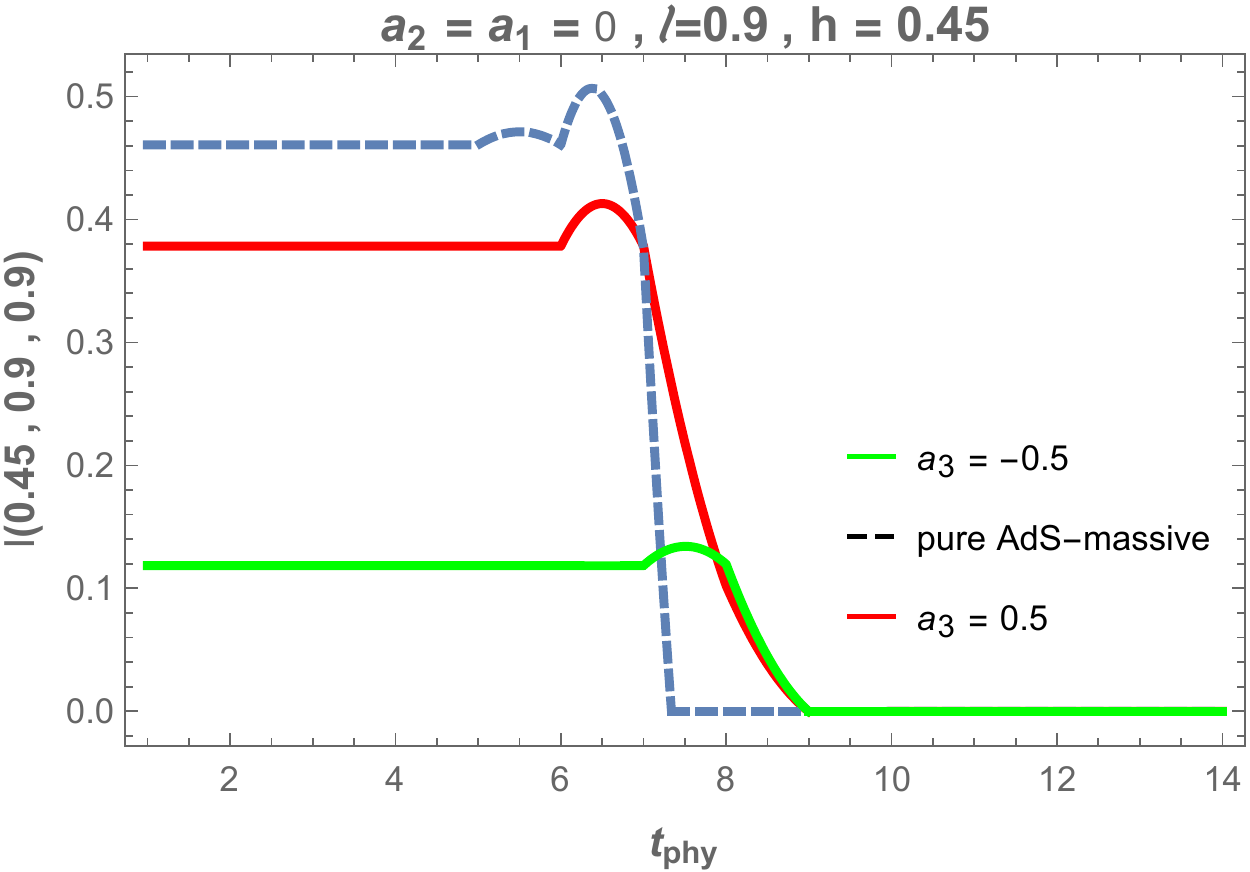}
		\includegraphics[scale=0.6]{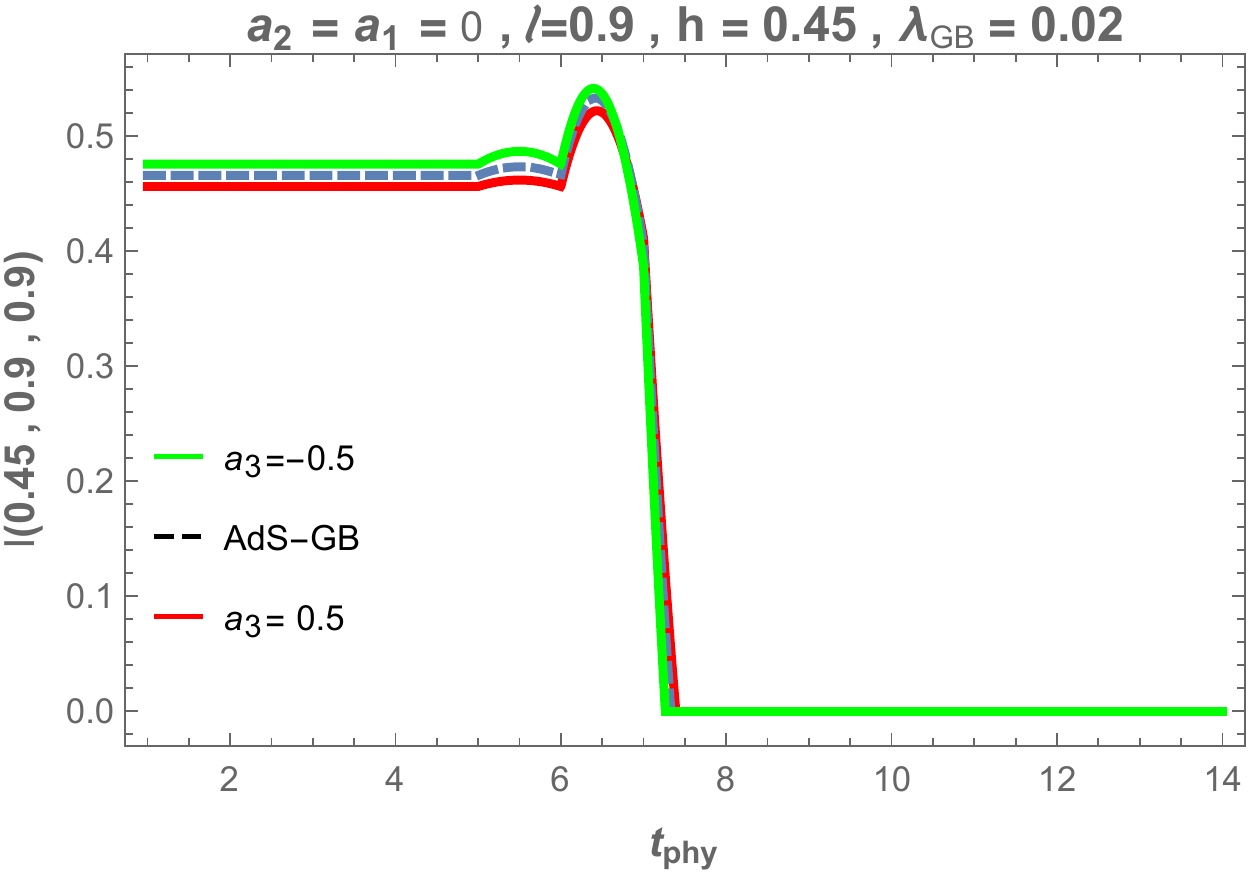}
	\caption{\it Time evolution of mutual information of  two disjoint intervals of  $\ell=0.9$ and distance separation $h=0.45$ in  ( Right panels) AdS-GB massive gravity and (left panels) AdS massive one in $d+1=5$ dimensional space-time. All graphs are obtained in the thin shell limit, $v_{s}=0.01$. The inset graphs shows the small changes in more details. }
	\label{fig:MI}
\end{figure}

In Fig \ref{fig:MI}, the effect of various massive parameters in the case of pure AdS and AdS-GB massive gravity with coupling $\lambda_{GB}=0.05$ on the time evolution of holographic mutual information is explored.
For more illumination, we first plotted the time evolution of mutual information for AdS massive gravity and investigated the effect of massive terms on pure AdS space-time then the contribution of GB coupling is considered. 
In general, there are four different regimes of time evolution of mutual information depending on the size of entangling regions or their separations. The logical  picture is as follows, when, say distance separation $h$ between two disjoint intervals, gets larger, the mutual information more decreased. 
In this section we do not discuss those topics. We examined the effect of massive terms and GB coupling on one of the regimes of time evolution of mutual information in $d+1=5$ dimensional bulk space-time. The other regimes have the same behavior. 
 In all plots, the mutual information starts from a non-zero value, then in two stages increases and decreases till to reaches its final maximum. After that, decreases to zero (depending on the choice of distance $h$ can be led to a non-zero value $I\neq 0$). It is worth noting that in $d+1=3,4$ dimensional space-time there is only one maximum point before the quench is applied~\cite{Allais:2011ys,Balasubramanian:2011at} (we checked this out for validity of our numerical code). Turning on the massive terms $a_i$'s in pure AdS space-time, one can observe that the behavior of $a_3$ is different of the others that is the $a_i>0$ where $i=1,2$  increases the mutual information in cases of pure AdS (dashed curve in left panels of Fig \ref{fig:MI}) but with different scales as shown while for $a_3$ both signs decrease the pure AdS mutual information.
Turning on the massive potentials in AdS-GB backgrounds in case of $a_2>0$ and $a_3>0$ can be led to the reduction of mutual information in pure AdS-GB massive background (dashed curves in the right panels of Fig \ref{fig:MI}).   
 The other observation is about the amount of decrease or increase of mutual information in presence of different massive terms. The $a_3$ term has more impact on the mutual information in a time-dependent AdS massive and AdS-GB massive backgrounds. For instance in AdS massive case, the more negative $a_3$ gets, the more mutual information decreases. In general, these changes in presence of GB corrections are smaller as it is observable clearly in the graphs of right panels of Fig \ref{fig:MI}.
\section{Conclusion}\label{conclusion}

%Here we have only made some observations about the effect of massive and Gauss-Bonnet terms on the holographic entanglement entropy and mutual information in a time-dependent background which is modeled by an AdS-Vaidya metric. We hope to explore the underlying physics of these observations in future works. 

In this paper, we studied numerically the effect of various massive potentials ($m^{2}c_{i}\,\mathcal{U}_i$) of dRGT massive gravity in pure AdS and AdS Gauss-Bonnet backgrounds on some of the parameters of  thermalization process such as final value of HEE (thermal value), the saturation time, the size of their kinks and finally on the mutual information. For sake of simplicity, the coefficients of the massive terms in the potential are denoted by $a_{i}=c_{i}c_{0}^{i}m^{2}$ and their effects on the thermalization were separately analyzed. The holographic approach to the problem is based on a time dependent background of the gravitational collapse of a thin shell of null dust to a black hole, the AdS-Vaidya geometry. 

Results indicated that for time evolution of HEE, the profiles of RT surfaces have the same pattern of an AdS background. The only difference appears in the final value of the thermal entropy. We observed an interesting and strange behavior for $a_1$ term when added to the pure AdS background for entangling intervals $\ell \leq 0.8$; Both signs of $a_1$ decrease the thermal value of entropy while for other terms ($a_2$ and $a_3$), the thermal entropy increases (decreases) when $a_2$ and $a_3$ are positive (negative) in both AdS massive and AdS-GB massive gravities. Beyond the above length, the strange behavior is related to the $a_3$ term.

In \cite{Blake:2013owa}, the graviton mass is related to the inhomogeneity of the dual background. It is logical that the more inhomogeneity of the boundary is, the faster thermalizing the system after a global quench. 

It is interesting to explore the role of each $a_i$ term of the massive potential in the saturation time individually to know which term can thermalize the system faster or in other words, which potential introduces more inhomogeneity on the boundary. For small entangling regions (i.e. $\ell \leq 0.8$), the saturation time is independent of the sign of $a_i$ where $i=1,2$ and above that length, the more positive $a_2$ is, the system thermalized faster. The $a_3$ term at sufficiently small lengths (i.e. $\ell \leq 0.4$) has different behavior. Here, $a_3>0$ first increases $t_{sat}$ then at some length, changes its behavior and decreases the saturation time as $a_2>0$ and $a_1>0$. This pattern is observed for either backgrounds mentioned above but this strange behavior is milder by considering the Gauss-Bonnet corrections. In summary, the system is thermalized faster in order of $t_{sat}(a_{1})>t_{sat}(a_{2})>t_{sat}(a_{3})$ at large entangling regions. As mentioned before there is a discrepancy between the saturation time calculated by the massless gravities and experimental measurements~\cite{Gyulassy:2004zy}. Our  results are important in the sense that it may resolve this discrepancy and bring theoretical values close to the observations.  

The next observation is related to the effect of these potentials on the kinks which appear in the evolution of HEE at large entangling regions. We observed the \textit{negative} potentials have the ability to reduce or \textit{eliminate} this swallow tails in the time evolution of HEE in pure AdS massive background. The order of this reduction in the size of the kink is as $|\Delta K(a_{3})|>|\Delta K(a_{2})|>|\Delta K(a_{1})|$. The positive $a_1$ and $a_2$ can produce or enlarge the existed kink and their orders in enlargement is as $a_{2}>a_{1}$. Both positive and negative $a_3$ significantly reduce the kink size. To explain why these happen a complete analytical approach is needed.

In an AdS background, turning on the positive (negative) massive potentials (except of $a_3$) can be led to increase (decrease) of mutual information. Both signs of $a_3$ reduces the mutual information but $a_3<0$ make notably decreases. 

Given an AdS-GB background, the positive (negative) potentials of $a_i$ where $i=2,3$ decrease (increase) the mutual information (the top insets of right panels of Fig \ref{fig:MI}). The $a_1$ behaves in the opposite way. The distinct difference in turning on the massive potentials between pure AdS and AdS-GB backgrounds is in the amount of change in the mutual information. In case of AdS massive gravity, these changes are significant but in the other one are negligible. 

These results are obtained by numerical methods. It would be very interesting to study the model analytically to achieve a better understanding of the nonlinear behavior of massive potentials in the thermalization process. We hope to address this issue in more details in our future works.

\end{document}